\documentclass[aps,12pt,preprint,superscriptaddress,showpacs]{revtex4}
\usepackage{amssymb,latexsym,amscd,amsmath,epsfig}
\usepackage{longtable}

\def\be{\begin{equation}}
\def\ee{\end{equation}}
\def\ba{\begin{array}}
\def\ea{\end{array}}
\def\bea{\begin{eqnarray}}
\def\eea{\end{eqnarray}}
\def\bi{\begin{itemize}}
\def\ei{\end{itemize}}
\def\bra#1{\langle #1|}
\def\ket#1{|#1\rangle}

\def\half{{\textstyle{1\over2}}}

\begin{document}
\title{Modeling Nuclear Pasta and the Transition to Uniform Nuclear Matter with the 3D Skyrme-Hartree-Fock Method at Finite Temperature I: Core-Collapse Supernovae}
\author{W.~G.~Newton}
\affiliation{Department of Physics, Texas A\&M University-Commerce, Commerce, Texas 75429-3011, USA}
\author{J.~R.~Stone}
\affiliation{Department of Physics,
University of Oxford, OX1 3PU Oxford, UK}
\affiliation{Department of Chemistry and Biochemistry, University of Maryland, College Park, MD 20742, USA}
\affiliation{Department of Physics and Astronomy, University of Tennessee, Knoxville, TN 37996, USA}
\date{\today}
\begin{abstract}

The first results of a new three-dimensional, finite temperature Skyrme-Hartree-Fock+BCS study of the properties of inhomogeneous nuclear matter at densities and temperatures leading to the transition to uniform nuclear matter are presented. Calculations are carried out in a cubic box representing a unit cell of the locally periodic structure of the matter. A constraint is placed on the two independent components of the quadrupole moment of the neutron density in order to investigate the dependence of the total energy-density of matter on the geometry of the nuclear structure in the unit cell.

This approach allows self-consistent modeling of effects such as (i) neutron drip, resulting in a neutron gas external to the nuclear structure, (ii) shell effects of bound and unbound nucleons, (iii) the variety of exotic nuclear shapes that emerge, collectively termed `nuclear pasta' and (iv) the dissolution of these structures into uniform nuclear matter as density and/or temperature increase. In part I of this work the calculation of the properties of inhomogeneous nuclear matter in the core collapse of massive stars is reported. Emphasis is on exploring the effects of the numerical method on the results obtained; notably, the influence of the finite cell size on the nuclear shapes and energy-density obtained. Results for nuclear matter in beta-equilibrium in cold neutrons stars are subject of part II. The calculation of the band structure of unbound neutrons in neutron star matter, yielding thermal conductivity, specific heat and entrainment parameters, will be outlined in part III.

Calculations are performed at baryon number densities of $n_{\rm b}$ = 0.04 - 0.12 fm$^{\rm -3}$, a proton fraction of $y_{\rm p}=0.3$ and temperatures in the range 0 - 7.5 MeV.  A wide variety of nuclear shapes are shown to emerge. It is suggested that thermodynamical properties change smoothly in the pasta regime up to the transition to uniform matter; at that transition, thermodynamic properties of the matter vary discontinuously, indicating a phase transition of first or second order. The calculations are carried out using the SkM$^*$ Skyrme parameterization; a comparison with calculations using Sly4 at $n_{\rm b}$ = 0.08 fm$^{\rm -3}$, $T$ = 0 MeV is made.
\end{abstract}

\maketitle

\section{\label{sec1}Introduction}
Observational properties of supernovae (SNe) and neutron stars (NSs) can serve as powerful constraints on the properties of bulk nuclear matter in density and temperature regimes inaccessible to laboratory experiments. In order that the theoretical description of nuclear matter be constrained by those properties, that description should be, as far possible, consistent over the whole extent of the relevant parameter space, a formidable task given that such matter spans fifteen orders of magnitude in density and reaches temperatures of up to $10^{12}$K in astrophysically relevant situations.

The main theoretical contribution to the description of nuclear matter comes in the form of the equation of state (EoS) of nuclear matter, obtained once the composition of the matter is specified. Such information is sufficient for the description of the bulk quasi-static properties of stars; however, in recent years increased sophistication in the modeling of collapsing stellar cores \cite{Liebendorfer2008}, of the cooling of NSs \cite{gnedin01,page04,Yak2008}, of oscillations at or near their surface \cite{Andersson2002,Samuelsson2007}, and of glitches in pulsar timing \cite{horva04,larso02,crawf03,jones98} have demanded a wider range of physical parameters to be theoretically determined for use in hydrodynamic and thermal transport calculations. Such nuclear matter properties as specific heat capacity, electrical and thermal conductivity, shear moduli, entrainment parameters and neutrino opacities should be calculated within the same theoretical framework and at the same level of approximation as the EoS~\cite{Lin2007}. It is to this end that we begin a self-consistent study of the properties of nuclear matter for use in stellar modeling.

The wide range of parameter space realized in SNe encompasses several possible phases of matter. Above nuclear saturation density $n_{\rm s} \approx 2.5 \times 10^{14}$ g cm$^{\rm -3}$ nuclear matter is uniform; at a density a little below nuclear saturation density nuclear matter becomes unstable to clustering \cite{peth95} and becomes inhomogeneous. The exact transition density has been shown to be sensitive to details of the nuclear matter EoS; at zero temperature, it is below $\sim 0.7 n_{\rm s}$ \cite{lattimer78,lattimer81,lattimer91}, and decreases with increasing temperature. Above a certain temperature $T \sim 10 - 20$ MeV, the inhomogeneous phase does not exist at all. At $T = 0$ MeV and below $\sim 0.2 n_{\rm s}$, the inhomogeneities are manifest as roughly spherical quasi-nuclei coexisting with a uniform fluid of neutrons, electrons (ensuring charge neutrality), and a degenerate gas of trapped neutrinos. Equilibrium with respect to weak interaction processes is not reached on the timescale of collapse and the proton fraction is roughly constant $\approx$ 0.3 \cite{Freedman74,Sato75}. As the density of nuclear matter increases, the separation of the quasi-nuclei becomes comparable with their physical extent: the scale on which the nuclear Coulomb interaction and surface tension act become similar. The matter becomes frustrated, characterized by a large number of local energy minima rather than a single distinct ground state. The transition to uniform matter through the density regime of frustrated matter is mediated by a series of more gradual transitions in the shape of the quasi-nuclei into more exotic structures: rods, slabs, cylindrical holes and bubbles, collectively termed `nuclear pasta'~\cite{rave83,hashi84,hashi84_2}. These structures have been determined on geometrical grounds by the competition between the nuclear surface tension and Coulomb repulsion of adjacent nuclear formations, and appear to be generic properties of frustrated matter; indeed one finds interesting parallels with structures observed in terrestrial complex fluids \cite{watanabe05}.

The pasta phases can contribute 10-20\% of the mass in the later stages of a collapsing stellar core~\cite{sonoda07}. As such, it is important to develop realistic, self-consistent models for this region of transition from inhomogeneous to uniform matter, and to determine the pasta phase diagram
and its dependence on nuclear matter properties (symmetry energy, compressibility) which are somewhat uncertain at sub-saturation densities and finite temperatures~\cite{sonoda08}.

In most theoretical models of inhomogeneous nuclear matter it is \emph{assumed} that at a given temperature and density the matter is arranged in a periodic structure throughout a sufficiently large spatial region for a unit cell to be identified. Given this, the properties of only one unit cell need be calculated. The basic properties of the pasta phase were first computed using semi-classical liquid drop models or Thomas-Fermi methods~\cite{lassaut87,lattimer91}. The energetically preferred pasta phase is obtained by calculating the energy-densities of the separate phases (including that of uniform matter) across the whole density and temperature regime in which pasta is expected to exist, and selecting the phase that gives the lowest energy-density at a given baryon number density. This means that the nuclear shapes expected to appear must be specified \emph{a priori}.

Fully microscopic one-dimensional (1D) Hartree-Fock (HF) calculations of SN matter were carried out by Bonche and Vautherin~\cite{bonche81,bonche82} (from hereon BV). These calculations did not include a systematic study of possible spurious effects, e.g. shell effects caused by the discretization of the unbound neutron energy spectrum by the finite computational domain.

While the 1D-HF method self-consistently describes the nuclear bulk, surface and shell effects, it restricts the nuclear shapes to being spherically symmetric. It requires the spherical Wigner-Seitz (WS) approximation, in which the generally non-spherical unit cell is replaced by a spherical one with the same volume. The WS approximation is good as long as the nuclear structures that form the lattice points are sufficiently widely spaced. This condition is satisfied everywhere except close to the transition density, when their spacings become comparable with their individual physical extent. Particularly, the transition to uniform matter cannot be treated in the WS approximation; the spherical unit cells by their nature do not continuously fill the space. In order to treat the transition to uniform matter correctly, three-dimensional (3D) studies are necessary.

The techniques of quantum molecular dynamics (QMD) and similar \cite{maruy1998} are semi-classical microscopic approaches that lead to an improved treatment of the pasta regime. In this method a large number of nucleons is dynamically evolved in a large cubic box with periodic boundary conditions and without assumption on the nuclear shape. The box can be made large enough to include the effects of electron screening on the Coulomb lattice energy. The pasta's dynamical response to a neutrino flux has been investigated in this framework \cite{horowitz2004_1, horowitz2004_2}, and significant strength at low energies from excitations of the internal degrees of freedom of the pasta has been found. This may be an important effect in the treatment of neutrino interactions with matter in SN simulations, especially concerning shock revival. The pasta shapes themselves and their sequence have also been studied in detail using QMD at zero and finite temperature \cite{watanabe01_2, watanabe_2_05}. It has been shown that more complicated shapes intermediate between the canonical pasta shapes may exist, and that pasta can also be formed by compression of a bcc lattice of spherical nuclei on a timescale much shorter than that of core collapse. The drawback of the QMD method is that the effective interaction used is very schematic and that the important microscopic shell effects are not included.

Recently, the three-dimensional (3D) HF method in a cubic box using the Sly4 Skyrme parameterization was applied to neutron star matter by Magierski and Heenen, and G\"{o}gelein and Muther \cite{magie02, gogel07}. Pasta shapes coexisting with the external neutron gas emerged naturally, and in addition other exotic shapes were found because no prior selection of the nuclear shapes expected was required. Magierski and Heenen examined the shell structure of the external neutron gas, which, it was posited, arises from the scattering of the free neutrons comprising the external neutron gas off the nuclear clusters. This effect has two related consequences: (i) the energy distribution of the free neutrons is discretized like that of the bound nucleons, forming a shell structure; (ii) the scattering causes an effective interaction between nuclear clusters analogous to the Casimir effect in quantum field theory (and led it to be dubbed the Fermionic Casimir effect) \cite{magie03}. The energy associated with that interaction is comparable to the energy difference between the different shape phases. As a result, the order in which the nuclear shape changes occur with changing density may be different to the sequence postulated in simpler models, and several shapes may coexist at the same density in different areas of the star, destroying the long range periodicity of the matter.

Both 3D-HF studies to date have calculated nuclear configurations at a limited number of values of densities and number of nucleons in the unit cell, and only for proton fractions expected to be found in neutron star crustal matter. Only G\"{o}gelein and Muther \cite{gogel07} performed calculations at finite temperature. To self-consistently probe the energy of various pasta shapes, both independent quadrupole moments ($q_{20}$ and $q_{22}$) should be constrained; the study of Magierski and Heenen \cite{magie02} imposed a constraint only on the $q_{20}$ component of the proton quadrupole moment, whereas G\"{o}gelein and Muther performed unconstrained calculations. Finally, it is not clear what role the numerical method itself plays in the results obtained; for example, what are the effects of the finite computational cell?

In the present work the first comprehensive 3D-HF study at finite temperature with a Skyrme-type interaction \cite{Skyrme1956,Vaut1972,Langa91,sto07} of the properties of inhomogeneous nuclear matter near the transition density to uniform matter in SN matter is presented. Preliminary results were outlined earlier \cite{new06}; in this paper a full account of the model is given including an extensive survey of the computational details necessary to prove validity of the approach to modeling nuclear pasta. Special attention is paid to the role played by the finite size of the computational cell. Furthermore, this work concentrates on a systematic exploration of the `shape phase space' of nuclear pasta using a strict constraint on the two independent quadrupole moments of the neutron density distribution to probe the energy of various nuclear configurations. The 3D-HF calculation across the phase transition region allows the EoS of nuclear matter to be developed self-consistently from neutron drip density (where a 1D-HF model is sufficient) right through to uniform matter densities, avoiding any artificial discontinuities that might occur matching together EoSs covering different density regions which has been a common practice so far.

The computational scheme presented in this paper is constructed to be flexible enough to easily adopt a variety of nuclear interactions other than those of the Skyrme type, and to study different boundary conditions. Its application to beta-equilibrium neutron star matter will be presented in part II of this work. Inclusion of more general Bloch boundary conditions and results of the first self-consistent calculation of the band structure of neutrons in the pasta phases of neutron star matter will be presented in the third paper (part III).

The paper is organized as follows: the theoretical method is described briefly in Sec.~\ref{sec2}. Implementation of the model in the form of the TAMAR code is outlined in Sec.~\ref{sec2a}. Numerical application to terrestrial nuclei and to inhomogeneous matter is detailed in Secs.~\ref{sec2b} and \ref{sec2c}. In Sec.~\ref{sec3} the results for supernova matter are presented and discussed. Conclusions and outline of future developments are given in Sec.~\ref{sec4}.

\section{\label{sec2}The Skyrme-Hartree-Fock (SHF) + BCS MODEL}
The computational framework used is briefly outlined in this section. Details relevant to the calculation of inhomogeneous nuclear matter in stars are given; more complete descriptions of the Hartree-Fock method are available elsewhere~\cite{Bonche1987, Bender2003, thesis}.

In the HF method, the ground state wavefunction is approximated by a single Slater determinant $\ket{\Phi} = \ket{ n_1, n_2,.....} = \hat{a}^{\dag}_1 \hat{a}^{\dag}_2 ..... \ket{0}$ which is found by minimization of the expectation value of the hamiltonian of the system $\delta \langle \Phi| \hat{H} | \Phi \rangle = \delta \mathcal{E}_{\rm Skyrme}[\Phi] = 0$, where $\mathcal{E}_{\rm Skyrme}$ is the energy-density functional \cite{sto07}. Minimization with respect to the single particle wavefunctions $\ket{n_i}$ yields a set of single-particle Schr\"odinger equation with the one-body HF potentials $u_q$: \be \label{2:E21a} \bigg[ - \nabla {\hbar^2 \over 2 m_q^*}
\nabla + u_q({\bf r}) + {\bf u}_{{(\rm so)},q}({\bf r}) \cdot {(\nabla \times \hat{\sigma}) \over
i} \bigg] \phi_{i,q}({\bf r}) = \epsilon_{i,q} \phi_{i,q} ({\bf r}) . \ee
\noindent Here, $q =$ p,n labels the isospin states, $i$ the single particle states, ${\bf u}_{{(\rm so)},q}$ is the spin-orbit potential, and $m_q^*$ is the effective mass.

Using the Skyrme interaction the one body potentials are obtained \cite{Vaut1972, Langa91}:
\begin{align} \label{2:E22}
u_q &= t_0 (1+ \half x_0) \rho - t_0(\half + x_0) \rho_q \notag
\\
       & \;\;\; + {1 \over 12} t_3 \rho^{\alpha} \bigg[ (2 +
       \alpha)(1 + \half x_3)\rho - 2(\half + x_3) \rho_q -
       \alpha(\half + x_3) {\rho^2_{\rm p} + \rho^2_{\rm n} \over \rho}
       \bigg] \notag \\
       & \;\;\; + {1 \over 4}[ t_1 (1 + \half x_1) + t_2 (1+ \half
       x_2)] \tau - {1 \over 4}[ t_1(\half + x_1) - t_2 (\half +
       x_2)] \tau_q \notag \\
       & \;\;\; - {1 \over 8}[3 t_1 ( 1 + \half x_1) - t_2(1 +
       \half x_2)] \nabla^2 \rho + {1 \over 8} [3 t_1 (\half + x_1) + t_2 (\half +
       x_2)] \nabla^2 \rho_q \notag \\
       & \;\;\; - \half t_4 (\nabla \cdot {\bf J} + \nabla \cdot {\bf J}_q) \; ,
\end{align}

\be \label{3:spin-orbit} {\bf u}_{{(\rm so)},q} = {1 \over 2} t_4 (\nabla \rho + \nabla \rho_q) + {1 \over 8} (t_1 - t_2) {\bf J}_q - {1 \over 8}(x_1 t_1 + x_2 t_2) {\bf J} ,\ee

\noindent and the effective mass is \begin{align} \label{2:E22a} {\hbar^2 \over 2 m_q^*} &= {\hbar^2 \over 2
m_q} + {1 \over 4}[ t_1 (1 + \half x_1) + t_2 (1 + \half
       x_2)] \rho - {1 \over 4}[ t_1(\half + x_1) - t_2 (\half +
       x_2)] \rho_q \; .
\end{align}
\noindent where $t_i$ and $x_i$ are parameters of the Skyrme interaction, $\rho = \rho_{\rm p} + \rho_{\rm n}$ are the nucleon densities, $\tau = \tau_{\rm p} + \tau_{\rm n}$ are the kinetic energy densities and ${\bf J} = {\bf J}_{\rm p} + {\bf J}_{\rm n}$ is the spin current. Two sets of Skyrme parameters were used in the present work; the SkM$^*$ parameterization~\cite{Bartel1982} based on a high precision description of nuclear ground states as well as surface energies and fission barriers of nuclei is used throughout the paper, and the Sly4 parameterization~\cite{Chaba1998} developed to describe neutron rich nuclei and neutron matter and biased more towards astrophysical applications.

In order to reduce the computational task in the present work the spin-orbit force has been not included in the HF Hamiltonian. Undoubtedly the spin-orbit interaction is an important component in determining the correct shell energy and single particle spectrum. However, the initial aim is to set up and thoroughly test the computational procedure, and to examine the optimum way to apply it to the calculation of the sub-nuclear EoS. The results are to be seen as a reference set with which later developments can be compared, and, in this case, the effect of the spin-orbit force systematically examined.

BCS pairing correlations are added to the Skyrme energy-density functional at zero temperature as outlined in \cite{thesis}. At finite temperature the single particle level occupation probabilities are given by the Fermi-Dirac distribution.

In the uniform matter case single particle wavefunctions can be represented by plane waves, the gradient terms in Eq.~(\ref{2:E22}) disappear and the expressions for the single particle potentials and the total energy-density of nuclear matter are analytic. A full explanation of obtaining the uniform matter EoS using the Skyrme-Hartree-Fock model can be found in~\cite{stone03}.

\subsection{\label{sec2a}Computational Implementation}
Eqs.~(\ref{2:E21a}) are solved in co-ordinate space in one cubic unit cell of the locally periodic nuclear matter. Note that this choice admits several different crystal lattice types; simple cubic (sc), body centered cubic (bcc) and face centered cubic (fcc). To reduce the computational task further, only nuclear configurations that conserve reflection symmetry in the three Cartesian directions are considered. It follows that the computation need be performed only in one octant of the unit cell.

It is expected that the absolute minimum energy of the cell is not going to be particularly pronounced and there will be a host of local minima separated by relatively small energy differences corresponding to different nuclear geometries. In order to systematically survey the `shape space' of all nuclear configurations of interest, the deformation of the nuclear configuration must be controlled via a constraint imposed on the nuclear density distribution. Reflection symmetry across the three Cartesian axes is assumed in the present model, thus eliminating asymmetric deformations such as dipole or octupole. A constraint is thus imposed on the quadrupole moment of the nuclear densities; the next order deformation consistent with the shape of the unit cell used, hexadecapole, is expected to give energy variations an order of magnitude smaller than those of the quadrupole deformation. The constraint is applied only to neutrons; because the matter that is modeled is neutron rich, the proton distribution is expected to follow the neutron distribution to a good degree of accuracy. The scheme used to implement the constraint can be found in~\cite{Cuss1985}. Specific deformations will be referred to in terms of the standard 'polar' deformation co-ordinates ($\beta, \gamma$) (see, e.g.~\cite{Bonche1985}).

The computational grid has a spacing $\Delta x$, $\Delta y$ and $\Delta z$ and is arranged so that the $i$th collocation point is located at $x_i = (i + \half) \Delta x$ and similarly for the $y$ and $z$ directions, where $i$ is an integer. The boundary conditions are simple periodic; a given function in the cell $\phi$ must obey $\phi ({\bf r} + {\bf T}) = \phi ({\bf r}).$ where ${\bf T}$ is the translation vector from the position ${\bf r}$ to the equivalent positions in the adjacent cells. These conditions are enforced by calculating derivatives and solving the Poisson equation for the Coulomb potential on the grid \cite{thesis} in Fourier space. The Coulomb solver utilises the FFTW software package \cite{fftw}. In NS matter, the more general Bloch boundary conditions must be used \cite{bloch,carter05}, but in SN matter (finite temperature and small density of dripped neutrons) this is not necessary. The Coulomb exchange term is evaluated via the Slater approximation~\cite{slater1951}. Integrals of functions are calculated using the trapezoidal rule; for periodic functions this is exact \cite{Baye1986}.

Two iteration methods were used in this work. The imaginary time step \cite{Davi1980} was used for the first tens of iterations for its robustness, and the damped gradient step \cite{Rein1982,Cuss1985} used afterwards for its quicker convergence at later iterations. The iteration is considered to have converged to the desired accuracy once the total variance in proton and neutron single particle energies $\langle \Delta h^2 \rangle_q = \sum_i w_{i,q} \bigg[ \bra{\phi_{i,q}} (h^n_{\rm HF})^2 \ket{\phi_{i,q}} - (\bra{\phi_{i,q}} h^n_{\rm HF} \ket{\phi_{i,q}} )^2 \bigg]$ becomes less than 1.0 keV$^{\rm 2}$ which ensures convergence in the total energy to less than one part in $10^8$.

A uniform background of electrons is included to ensure charge neutrality. The electrons generate their own Coulomb potential, $\Phi_e$ which is determined in the same way as the proton Coulomb potential. Screening of the electron Coulomb potential is taken into account simply by omitting the infinite part of the potential. Free Fermi gas expressions for the energy and entropy of the electrons are used~\cite{thesis}. The electrostatic interaction between the electrons and the inhomogeneous proton distribution is taken into account by adding to the single particle kinetic energies of the electrons an electrostatic potential term as in \cite{bonche81}.

The total energy of the nuclear configuration including the contribution of the Coulomb force from the electrons and protons is given by \cite{Langa91}: \be \label{energy1}
\mathcal{E}_{\rm skyrme} + \mathcal{E}_{\rm coul} = \half \bigg( \mathcal{E}_{\rm kin} + \sum_{\beta} w_{\beta} \epsilon_{\beta} \bigg) + \mathcal{E}_{\rm rearr} + \mathcal{E}_{\rm quad} + \mathcal{E}_{\rm pair} + \mathcal{E}_{\rm exch,coul} + \mathcal{E}_{\rm e,coul} + \mathcal{E}_{\rm lattice},\ee

\noindent where the various contributions are, respectively, from the kinetic energy of the nucleons, the single particle energies, the Skyrme rearrangement energy, the energy due to the quadrupole constraint, the pairing energy, the Coulomb exchange energy, the Coulomb interaction of the electrons and the lattice energy.

At finite temperature the Helmholtz free energy $F = \mathcal{E}_{\rm tot} - TS$ is calculated. Here $T$ is the temperature and $S$ is the total entropy of all particles present calculated using the free Fermi gas expressions derived in Appendix A of Ref.~\cite{thesis}. The free energy-density is then obtained by dividing by the volume of the unit cell $f = F/V$. The pressure density of the matter is calculated using $p = -f + \sum_i \mu_i n_i$ where the sum is over all particle species (labeled $i$), $\mu_i$ are the chemical potentials and $n_i$ the number densities. The evaluation of the chemical potentials is given in detail in~\cite{thesis}.

As a final note, in the following the \emph{mean} number densities of particles in the matter (that is, their bulk values) are simply denoted $n_{\rm n}, n_{\rm p}$, etc. Where necessary, the local number densities will be given as a function of position in the unit cell $n_{\rm n} (x,y,z)$, etc.

\subsection{\label{sec2b} Calculation of terrestrial nuclei}
The calculation scheme described above has formed the basis of the TAMAR computer code. The code has been extensively tested \cite{thesis} for terrestrial nuclei against two reference HF codes. The first is a 2D code SKYAX in cylindrical polar co-ordinates which enforces axial symmetry \cite{Umar1989}. This code uses a finite difference scheme to treat the derivatives and calculates the Coulomb potentials using a separate iterative process. The second is a three-dimensional code TDHF3D \cite{rein_private2}. Like the TAMAR code, it represents derivatives and solves the Coulomb force using Fourier Transforms, allowing for a more direct comparison. Both reference codes use the damped gradient iteration step. Since the three codes use different prescriptions for the calculation of the Coulomb and pairing forces, the test has been performed with the basic Skyrme nuclear interaction without the constraints or the Coulomb and pairing forces. Two Skyrme parameterizations SkM$^*$ and Sly4 have been used to calculate the binding energies and single particle energies of a selection of doubly magic nuclei and $^{56}$Fe and excellent agreement with both reference codes was found. As an example, in Table~\ref{Tab:1} the binding energies and root mean square proton and neutron radii as calculated by the three codes using the Sly4 Skyrme parameterization are compared.

Furthermore, the method used in the TAMAR code to calculate the Coulomb potentials has been carefully examined. This method is intrinsically periodic; when solving for a nucleus the presence of a surrounding simple cubic lattice of the same nuclei is implicitly taken into account. There is thus a contribution to the binding energy of the nucleus from the lattice. The lattice energy decreases with increasing the computational volume; for large volumes, effects of the finite size of the nucleus can be neglected and the lattice energy is given by that of a simple cubic lattice of point like nuclei of charge $Ze$: $E_L = -1.444 {Z^2 e^2 \over a^3}$. Testing of the  variation in total energy with cell size revealed that the Coulomb energy  tends towards a constant value at large $a$ as expected, and the contribution of the lattice energy falls off as $1/a^3$. The proton Coulomb energy was found to be in agreement with the TDHF3D reference code.

\subsection{\label{sec2c}Inhomogeneous matter}
\subsubsection{\label{sec2c1}General Numerical Considerations}
For each run of the code at a given temperature $T$ and baryon number density $n_{\rm b}$ = $A/V$, where $A = N + Z$ is the number of nucleons ($N$ neutrons and $Z$ protons), and $V$ the volume of the unit cell, up to four free parameters need to be specified: $A$, the proton fraction $y_{\rm p}$ = $Z/A$, and the quadrupole moment parameters $\beta$ and $\gamma$. Clearly at a given value of $n_{\rm b}$, $A$ and $V$ are not uniquely determined and calculations have to be performed over a range of values of $A$, or equivalently $V$, at that given density to find the value which gives the minimum free energy-density. The proton fraction is kept constant throughout this work at $y_{\rm p}=0.3$, a value appropriate for collapsing stellar cores \cite{lattimer91}. If the quadrupole constraint is applied, the deformation parameters $\beta$ and $\gamma$ are also free parameters and must be varied over the deformation space, searching for shapes corresponding to minima of the total energy.

To build an EoS of the pasta phases of inhomogeneous matter, such a series of calculations has to be performed at each point of the parameter space ($n_{\rm b}, T, y_p$) with adequate spacing, which is extremely computationally intensive. Part of the task will be to find the optimum strategy to obtain adequate coverage of parameter space; for example, some areas of parameter space are physically less important than others; this will lead to selection of those values of the parameters that are most likely to be physically manifest, and a corresponding reduction in the computation time. Examples of this selection process are given in Sec.~\ref{sec3}.

The calculation of inhomogeneous nuclear matter requires consideration of some numerical effects that are common to the calculation of isolated terrestrial nuclei and, in addition, a number of numerical effects specific to the particular problem at hand. The computational volume $V$ is uniquely defined by the baryon number density $n_{\rm b}$ and the number of nucleons $A$ present in that volume by $V$ = $A$/$n_{\rm b}$. Since $V$ is fixed at a given $n_{\rm b}, V$, the number of grid points and the grid spacing are inversely related - a constraint not present in the calculation of isolated nuclei. There is a balance between the number of grid points used to obtain the desired accuracy and the required computational time. The number of grid points determines the maximum number of orthogonal wavefunctions that can be represented on the grid. For the grid parameter of interest in this work ($\Delta x_{\alpha} \sim$ 1 fm, $n_{x_{\alpha}} \sim 10$), this number is in the thousands. Ideally the calculation should be performed using all the available wavefunctions as it is not known \emph{a priori} which is the preferred set of occupied states forming the configuration corresponding to the minimum total energy, and yielding a realistic approximation to the physical ground state. Furthermore, with increasing temperature a larger number of single particle states will become occupied, so a smaller value of grid spacing will be required for higher temperatures. However, in the interests of reducing the computation time, the number of wavefunctions participating in the iteration process should be minimal. As a compromise, each calculation begins by including all the wavefunctions available for a given grid; then, after a certain number of iterations, those wavefunctions which have occupation amplitude below $10^{-6}$ are excluded from the iteration process after each iteration. It was found that 100 iterations was the safe choice of the cut-off point.

The optimal grid spacing was determined as follows. Calculations were performed at a proton fraction of $y_{\rm p}$ = 0.3, baryon number densities of $n_{\rm b}$ = 0.06, 0.08 and 0.10 fm$^{\rm -3}$, temperatures T=0, 2.5, 5.0 and 7.5 MeV, quadrupole deformations $\gamma = 0^{\rm o}$ and $\beta$ = 0.0 and 1.0,  and nucleon number $A$ = 100-1200 with a step of 20. For each of these collections of points in parameter space, an average grid spacing of $\langle {\Delta}x_{\alpha} \rangle$ = 1.0, 1.1, 1.2 and 1.3 fm was tested \cite{thesis}, and the difference in energy obtained at each grid spacing was calculated. It was found that using a grid spacing of 1.0 fm guarantees an accuracy 1 part in $10^{-4}$ in the energy for temperatures up to 5 MeV. At 7.5 MeV, a smaller grid spacing is required. The calculations presented in this paper do not use grid spacings below 1.0 fm at this temperature because of the large computation time required, so care has to be taken in interpreting the results at this high a temperature, where, in principle, higher momentum states need to be included.

Another important requirement the model has to satisfy is the independence of the final outcome on the form of the initial wavefunctions as mentioned in Sec.~\ref{sec2a3}. Two cases were considered: (i) both proton and neutron wavefunctions are of the form Gaussian times polynomial (to assure orthogonality), (ii) the proton wavefunctions are Gaussian and the neutron wavefunctions are plane waves. It was found that both cases converged to the same energy to less than 1 part in 10$^{-4}$.

\subsubsection{\label{sec2c2}Effect of Finite Cell Size on the Computation of Inhomogeneous Matter}
When performing calculations of isolated nuclei the computational volume can be made much larger than the size of the nucleus, and is constrained only by the computation time. It follows that the effect of the choice of the computational volume can be minimized. In contrast, the nuclear configurations in stars span the entire computational volume; indeed the volume itself acquires a physical meaning. Thus the numerical effects of the size of a finite cell must be carefully examined.

Given a particular nuclear shape obtained in a cubic cell, it should in principle be possible to double the cell size and, searching over deformation space, find a configuration equivalent to two instances of the original shape adjacent to each other. The free energy-density obtained in the doubled cell should be identical to that in a single cell.

To examine this effect, calculations were performed in a box that is double the length in the z-direction compared to the other two directions. The expectation was to find a configuration similar to the one in a cubic box at the same baryon number density (with half the number of nucleons), with the same free energy energy.
Fig.~\ref{Fig:1} displays one of such configurations. The nucleon density distribution has been rendered in three dimensions, with blue indicating the lowest densities and red the highest. The right-hand cell displays the nuclear shape obtained within a cubic cell; it is similar to the `lasagna' phase, with an additional cylindrical bridge joining the slabs. The left-hand cell displays a shape that is very similar to two instances of the modified lasagna shape adjacent to each other. The free energy densities of the two configurations agree to within 1 part in $10^{\rm 4}$.

\begin{figure}[!b]
\hspace{1pc}
\centerline{\psfig{file=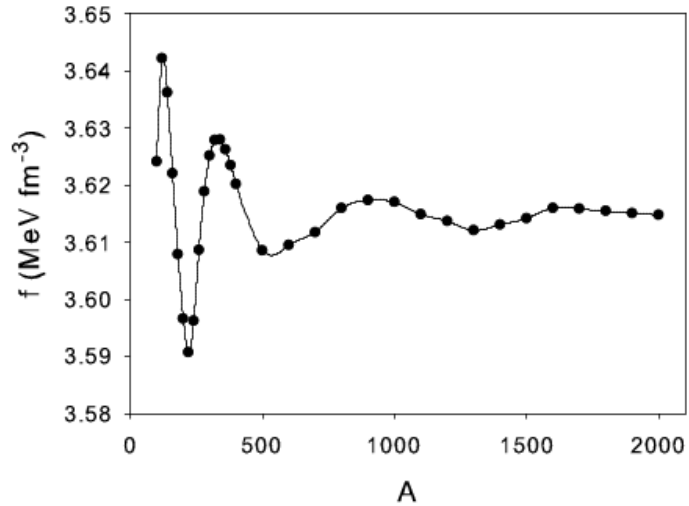,width=10cm}}
\caption{Free energy-density $f$ versus nucleon number $A$ at $n_{\rm b}$ = 0.11 fm$^{\rm -3}$, ($\beta,\gamma$)=(0,0$^{\rm o}$) and T = 2.5 MeV. The form of the curve is dominated by a spurious shell energy caused by the finite box discretization of the continuum neutron energy spectrum.} \hspace{1pc} \label{Fig:2}
\end{figure}

Fig.~\ref{Fig:1} demonstrates clearly that the nuclear structures obtained are not artifacts of the finite cell size (and more evidence will be presented in paper II for neutron star matter). However, the finite cell size can be a source of spurious shell effects. In analogy with a Fermi gas in a box, shell effects caused by the discretization of the physical space due to the finite computational volume may occur. These effects will manifest themselves especially at high densities and temperatures when a large number the nucleons are unbound. To illustrate the form of these numerical shell effects, the free energy-density is plotted in Fig.~\ref{Fig:2} as a function of $A$ in a wide region up to $A$=2000 at a density of $n_{\rm b}$ = 0.11 fm$^{\rm -3}$, $(\beta, \gamma)$ = (0.0,0$^{\rm o}$) and $T$ = 2.5 MeV. The pronounced oscillations in the free energy-density curve are entirely numerical in origin. With increasing box size and temperature the bulk energy-density of the matter in the box approaches that derived from the semi-analytic formulae for the free energy-density of a free nucleon gas. These numerically induced shell effects manifest themselves in a form distinct from the physical shell effects, arising due to a combination of the shell energies of bound nucleons and unbound neutrons scattered by the bound nucleons, which are characterized by more rapid fluctuations in nucleon number $A$ typically at lower densities and temperatures and lower values of $A$. The distinction between the form and occurrence of the two types of shell effects is encouraging as it allows their easy identification. In such a situation where the shell effects are purely spurious, the physical value of the free energy-density is not the minimum, but that value to which the free energy tends at high $A$.

In order to accurately locate free energy-density minima as a function of $A$ and in energy-density surfaces, the numerical shell effects, which introduce spurious minima, should be identified and corrected for.

\section{\label{sec3}Results and Discussion}

Keeping the proton fraction fixed at $y_{\rm p}$ = 0.3, and at selected densities and temperatures $n_{\rm b}$ and $T$, the variation of free energy-density of inhomogeneous nuclear matter $f$ with the parameters $A$, $\beta$ and $\gamma$ is now explored.

\subsection{\label{sec3ss1}Variation of free energy-density with nucleon number at constant deformation}
\begin{figure}[!t]
\centerline{\psfig{file=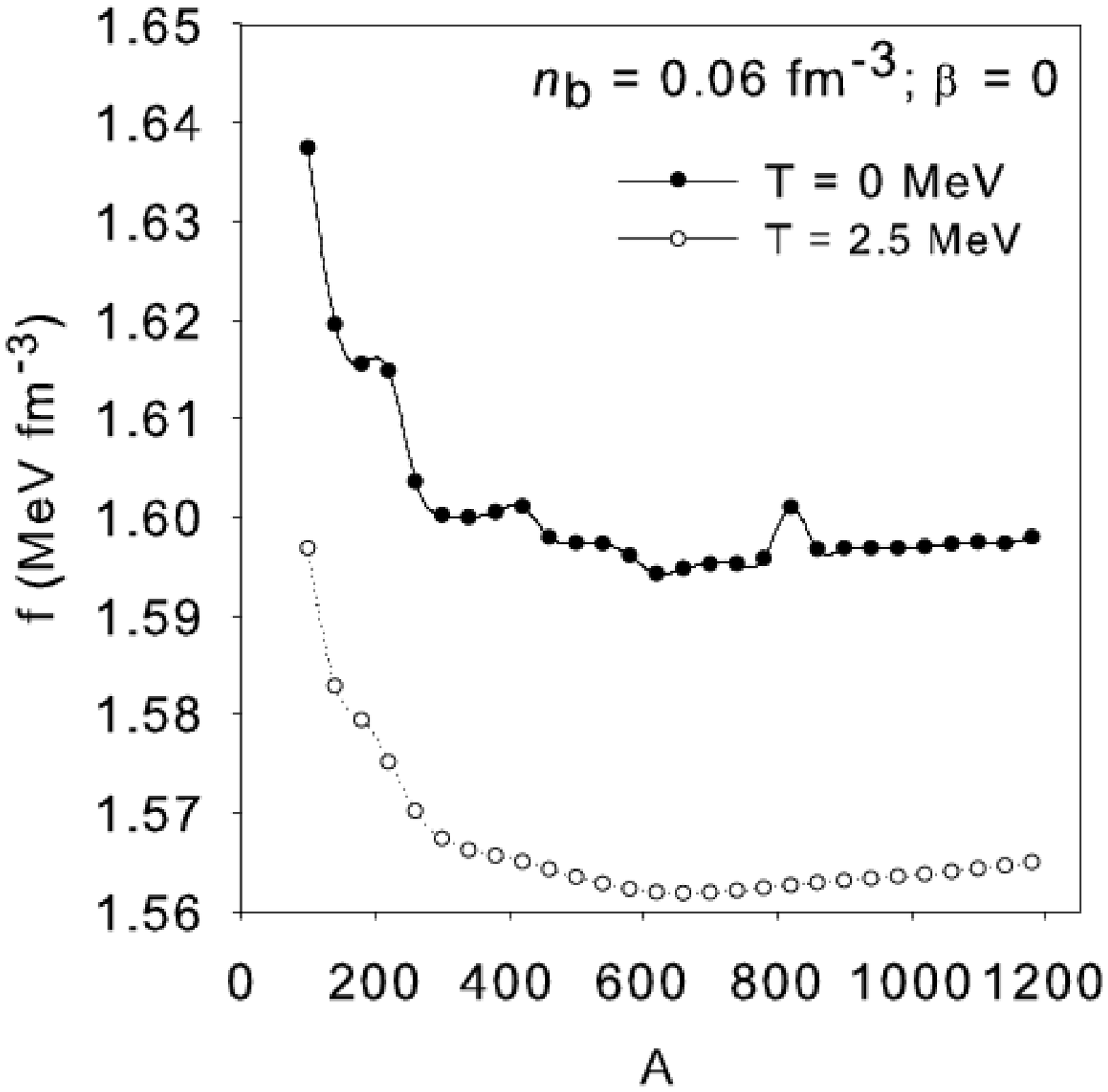,width = 8cm}\psfig{file=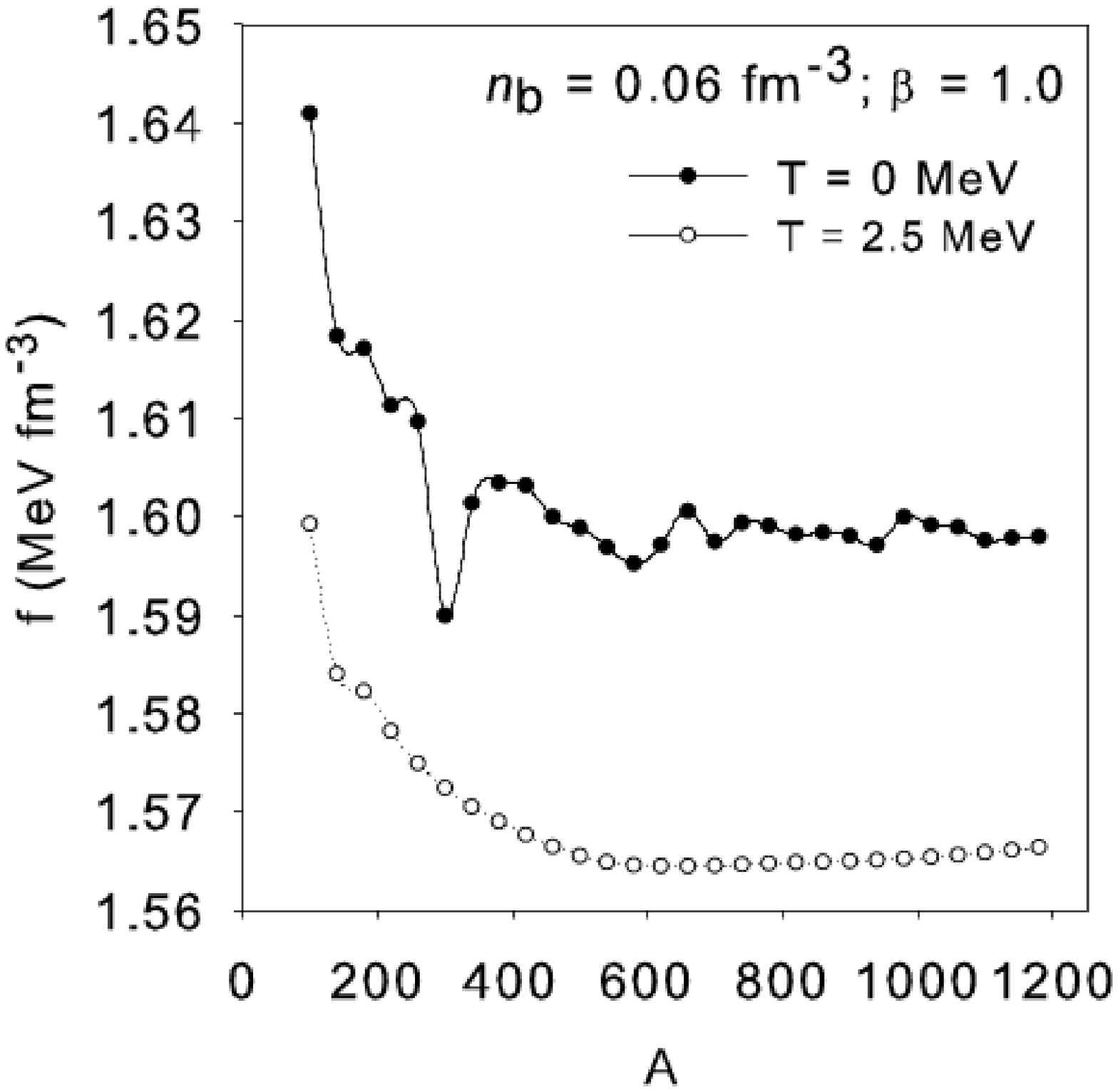,width = 8cm}}
\centerline{\psfig{file=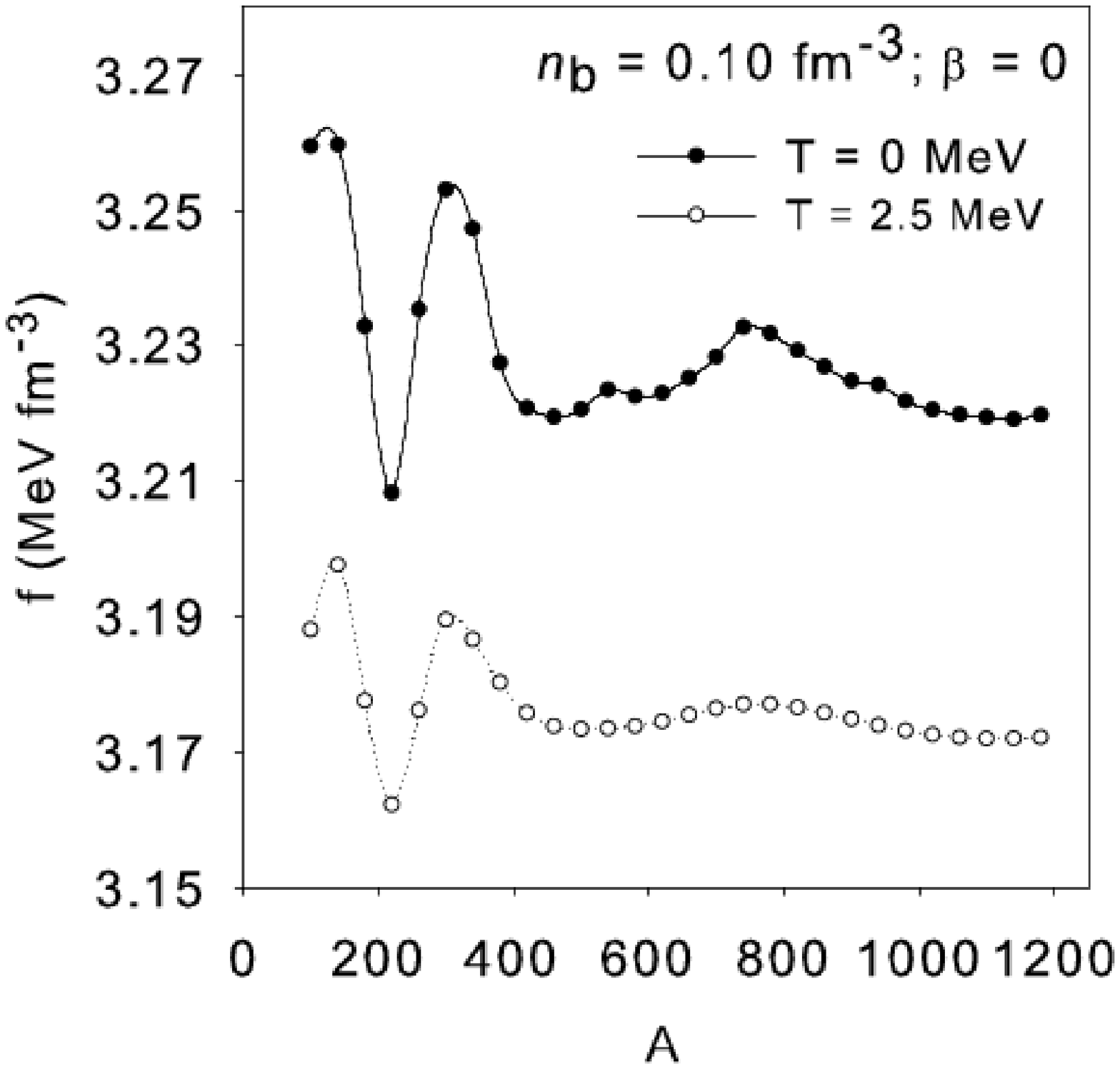,width = 8cm}\psfig{file=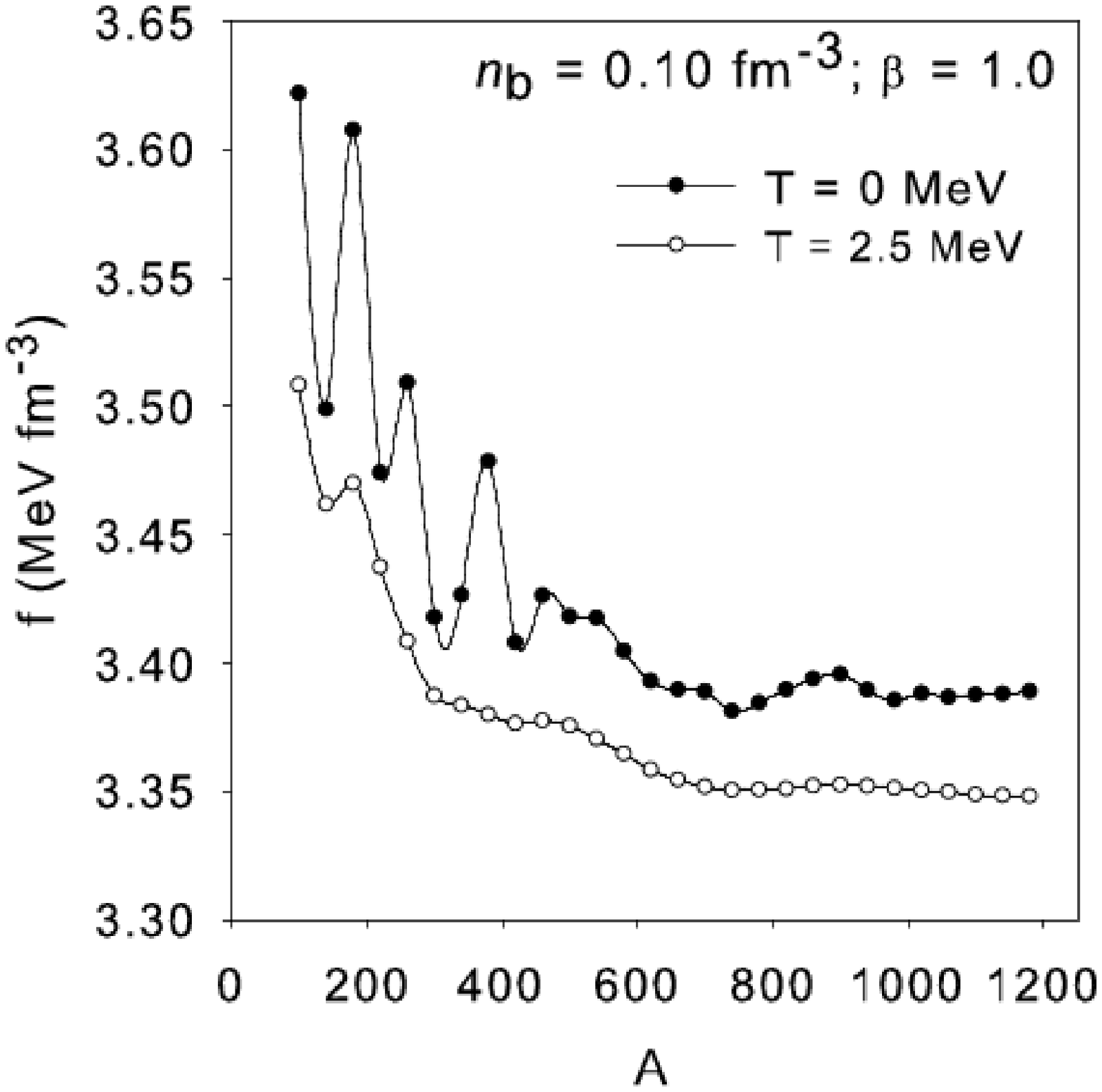,width = 8cm}}
\caption{Free energy-density $f$ as a function of the number of nucleons $A$ in the unit cell at different densities, temperatures and deformations. Top left: ($\beta, \gamma$) = (0.0, 0$^{\rm o}$), $n_{\rm b} = 0.06$ fm$^{-3}$. Top right: ($\beta, \gamma$) = (1.0, 0$^{\rm o}$), $n_{\rm b} = 0.06$ fm$^{-3}$. Bottom left: ($\beta, \gamma$) = (0.0, 0$^{\rm o}$), $n_{\rm b} = 0.10$ fm$^{-3}$. Bottom right: ($\beta, \gamma$) = (1.0, 0$^{\rm o}$), $n_{\rm b} = 0.10$ fm$^{-3}$. Results at $T$ = 0.0 MeV and $T$ = 2.5 MeV are shown on each plot.} \hspace{1pc} \label{Fig:3}
\end{figure}

\begin{figure}[!b]
\centerline{\psfig{file=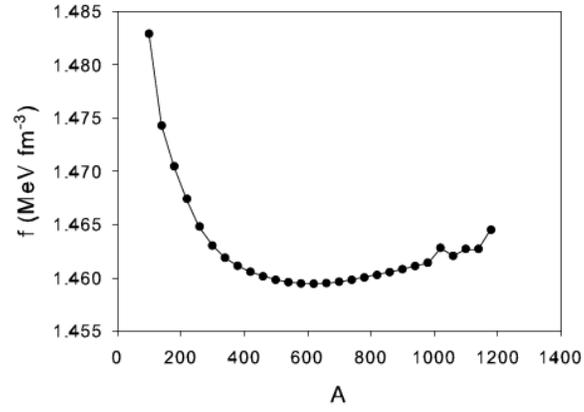,width = 8cm}}
\caption{Free energy-density $f$ as a function of the number of nucleons $A$ in the unit cell at $n_{\rm b} = 0.06$ fm$^{-3}$, ($\beta, \gamma$) = (0.0, 0$^{\rm o}$), $T$ = 5 MeV.} \hspace{1pc} \label{Fig:4}
\end{figure}

The minimization of free energy-density with respect to total nucleon number $A$ (equivalent to unit cell size) at constant deformation is illustrated in Fig.~\ref{Fig:3} at two densities $n_{\rm b}$ = 0.06 fm$^{\rm -3}$ (top two plots) and 0.10 fm$^{\rm -3}$ (bottom two plots). The deformation is held constant at $(\beta, \gamma) = (0.0, 0^{\rm o})$ for the left two plots and $(\beta, \gamma) = (1.0, 0^{\rm o})$ for the right two plots. Each plot displays results obtained $T$ = 0 MeV and $T$ = 2.5 MeV to demonstrate the effect of temperature on the form of the curves. Note that in general, the lower the temperature, the higher the free energy-density: increasing the temperature means the amount of useful energy (i.e. available for mechanical work) decreases as the entropy increases. In addition, the free energy-density versus $A$ at $n_{\rm b}$ = 0.06 fm$^{\rm -3}$, $T$ = 5 MeV, and $(\beta, \gamma) = (0.0, 0^{\rm o})$ is shown in Fig.~\ref{Fig:4}. The variation of important EoS quantities (free energy-density $f$, entropy density $s$ and pressure density $p$) with $A$ is tabulated in Table~\ref{Tab:2} for $n_{\rm b}$ = 0.06 fm$^{\rm -3}$, $T$ = 2.5 MeV and $(\beta, \gamma) = (1.0, 0^{\rm o})$.

The form of the free energy curves displayed is effectively a superposition of physical effects related to the geometry of the nucleon density distribution, physical shell effects and numerical shell effects. There are also occasional discontinuities in the curves (e.g. at $A$ = 820 in the plot at $n_{\rm b}$ = 0.06 fm$^{\rm -3}$, $T$ = 0 MeV and $(\beta, \gamma) = (0.0, 0^{\rm o})$) caused by a sudden change in the geometrical configuration of the nucleons in the unit cell, as will be discussed below.

The competition between nuclear surface tension and Coulomb energy manifests itself as a broad minimum. Because such minima have their origin in the nuclear geometry, they will be referred to from now on as geometrical minima. It is upon this broad curve that the shell effects are superimposed. The physical shell effects are particularly evident at zero temperature as irregular oscillations in energy which are responsible for local minima appearing in the curves. As noted in \cite{bonche81}, in matter with a proton fraction of 0.3, shell effects yield minima far from the familiar magic numbers of terrestrial nuclei. Shell effects have their highest magnitude at low values of $A$. This is because the density of states is lowest at smallest $A$; the energy gaps between shells are at their largest, thus the difference in total energy between a filled shell and, say, a shell occupied by only two nucleons, is relatively large. With increasing $A$, the density of states increases, the shell separations become smaller, and the effects on total energy diminish.

As the temperature increases, the shell effects become less pronounced, disappearing when the typical spacing between shells is less than $k_{\rm B} T$. This is particularly evident as one follows the form of the free energy curve at $n_{\rm b}$ = 0.06 fm$^{\rm -3}$, $(\beta, \gamma) = (0.0, 0^{\rm o})$ as one increases temperature from $T$ = 0 MeV through $T$ = 2.5 MeV to $T$ = 5 MeV (top left plot of Fig.~\ref{Fig:3} and Fig.~\ref{Fig:4}). At $T$ = 5 MeV the shell effects have completely disappeared and only the broad geometrical minimum remains at $A$ = 600. The plots of free energy-density with increasing $A$ at $n_{\rm b}$ = 0.06 fm$^{\rm -3}$ and zero deformation is consistent that found in the 1D-HF calculations of Bonche and Vautherin \cite{bonche81}. The absence of the spin-orbit interaction alters the details of shell effects but does not change the picture qualitatively. The broad minimum appears at $A$ = 600 in both works.

As deformation increases to $(\beta, \gamma) = (1.0, 0^{\rm o})$ at the lower density $n_{\rm b}$ = 0.06 fm$^{\rm -3}$, the curves remain qualitatively similar. The shell effects at $T$ = 0 MeV have increased in magnitude a little; indeed, the minimum appears at $A$ = 300 due to shell effects rather than the nuclear geometry. The geometrical minimum occurs at around the same value of $A$ despite the different nuclear geometry, although this is not always the case. The geometry at $(\beta, \gamma) = (1.0, 0^{\rm o})$ is displayed in Fig.~\ref{Fig:4}, and the geometry at $(\beta, \gamma) = (0.0, 0^{\rm o})$ in the top left panel of Fig.~\ref{Fig:12}. The free energy-density at the geometrical minimum is about 2 keV fm$^{-3}$ greater at $(\beta, \gamma) = (1.0, 0^{\rm o})$ than at zero deformation. Several EoS quantities are tabulated in Table~\ref{Tab:2} as they vary with $A$ at $n_{\rm b}$ = 0.06 fm$^{\rm -3}$, $T$ = 2.5 MeV, $(\beta, \gamma) = (1.0, 0^{\rm o})$. One can see that although the free energy changes relatively smoothly with $A$ at this temperature as most of the shell effects have been washed out, quantities such as pressure and entropy which depend on the derivative of the free energy still fluctuate a significant amount as $A$ increases.

The bottom two plots of Fig.~\ref{Fig:3} show the variation of free energy-density with $A$ at a higher density $n_{\rm b}$ = 0.10 fm$^{\rm -3}$. At zero deformation the form of the curves is qualitatively different from those at the lower density. The nucleon distribution is close to being uniform at this density (in fact there are small bubbles present in otherwise uniform nuclear matter; see the top two panels of Fig.~\ref{Fig:13}) and the shell effects are dominated by those of a uniform Fermi gas in a cubic box as discussed in Sec.~\ref{sec2c2} and illustrated for completely uniform matter in Fig.~\ref{Fig:2}. As has been discussed, these shell effects are numerical in origin and show a characteristic form significantly different to physical shell effects in bound nuclear structures; they are more regular and persist to higher values of $A$ and $T$. At this density, since structure is still present, they are overlaid on a broad geometrical minimum. The spurious shell effects give local minima deeper than the geometrical minima, so until they are subtracted off, it is no longer good practise to take the ground state of matter to be that corresponding to the minimum energy configuration. At zero temperature there are still some physical shell effects visible; these become washed out at $T$ = 2.5 MeV leaving behind the purely spurious effects superimposed on the geometrical minimum.

At $n_{\rm b}$ = 0.10 fm$^{-3}$, the free energy-density of significantly deformed matter ($\beta$ = 1.0) is appreciably higher than that of non-deformed matter. This is an indication that the transition to uniform matter is being approached: more uniform configurations are energetically preferred rather than highly deformed ones. The variation of $f$ with $A$ has a form similar to that at $n_{\rm b}$ = 0.10 fm$^{-3}$; the matter, being deformed, no longer exhibits the spurious shell effects associated with uniform nuclear matter.

3D renderings of a selection of neutron density distributions taken at $T = 0.0$ MeV and $T$ = 5.0 MeV for the sequences of free energy-density versus $A$ at $n_{\rm b}$ = 0.06 fm$^{\rm -3}$, ($\beta,\gamma$) = (0.0, 0$^{\rm o}$) (see top left panel of Fig.~\ref{Fig:3} and Fig.~\ref{Fig:4}) are shown in Fig.~\ref{Fig:5}. The unit cells are displayed to scale relative to each other, so the increase in cell size as $A$ increases is clearly visible. The smallest cell width is 11.9 fm and the largest is 27.1 fm. It can be seen that the constraint on both quadrupole moments has ensured that, \emph{for the most part}, the same geometry is obtained for each value of $A$, at constant ($\beta$, $\gamma$). Thus a unique geometrical configuration could be followed as the total nucleon number $A$ is increasing. The geometry also remains the same for the two temperatures shown. At zero temperature there is no neutron gas external to the nuclear structure, although a neutron halo is visible; at $T$ = 5.0 MeV an external neutron gas exists. The main structure obtained is notably not a member of the canonical pasta collection: its shape is similar to that of three cylinders intersecting each other at right angles. The proton distribution, not displayed, follows the neutron distribution very closely. The external neutron gas is very dilute at both temperatures, as has been found in previous studies of SN matter \cite{bonche81}.

Although for the most part, the same geometrical configuration is obtained at a given ($\beta$, $\gamma$), there are some configurations that do not follow the pattern (see the bottom panels). This is a result of the freedom inherent in the TAMAR code from not constraining higher-order deformations; for example, a configuration may have the same quadrupole moment, but a different hexadecapole moment. These departures from the energetically preferred geometrical configuration are rare, however, and are clearly distinguished from others in the energy plots as discontinuities in the curves of order of 1 - 5 keV fm$^{\rm -3}$. One such point is observed at $T = 0.0$ MeV at a nucleon number of $A$ = 820 and three at $T$ = 5.0 MeV: $A$ = 1080, 1140, 1200. The former has a small number of nucleons split off from the main structure, forming a `mini-nucleus' arranged in a body-centered cubic lattice with respect to the larger structure. The latter two points shows a more significant change: now the nucleons have arranged themselves more equally in a body-centered cubic structure with nucleon `arms' joining the nuclei at the lattice points.

\subsection{Variation of free energy-density with nucleon number and deformation parameters $\beta$, $\gamma$}

In the previous section the variation of free energy-density $f$ with nucleon number was considered at two specific values of the deformation parameters. In this section the study of the variation of $f$ is extended to include the whole range of values of $\beta$ of interest and for selected values of $\gamma$. This will give a series of energy-deformation surfaces $f = f(A,\beta,\gamma = const.)$ which contain information about shell and geometrical effects associated with varying nuclear shapes.

\begin{figure}[!t]
\hspace{1pc}
\centerline{\psfig{file=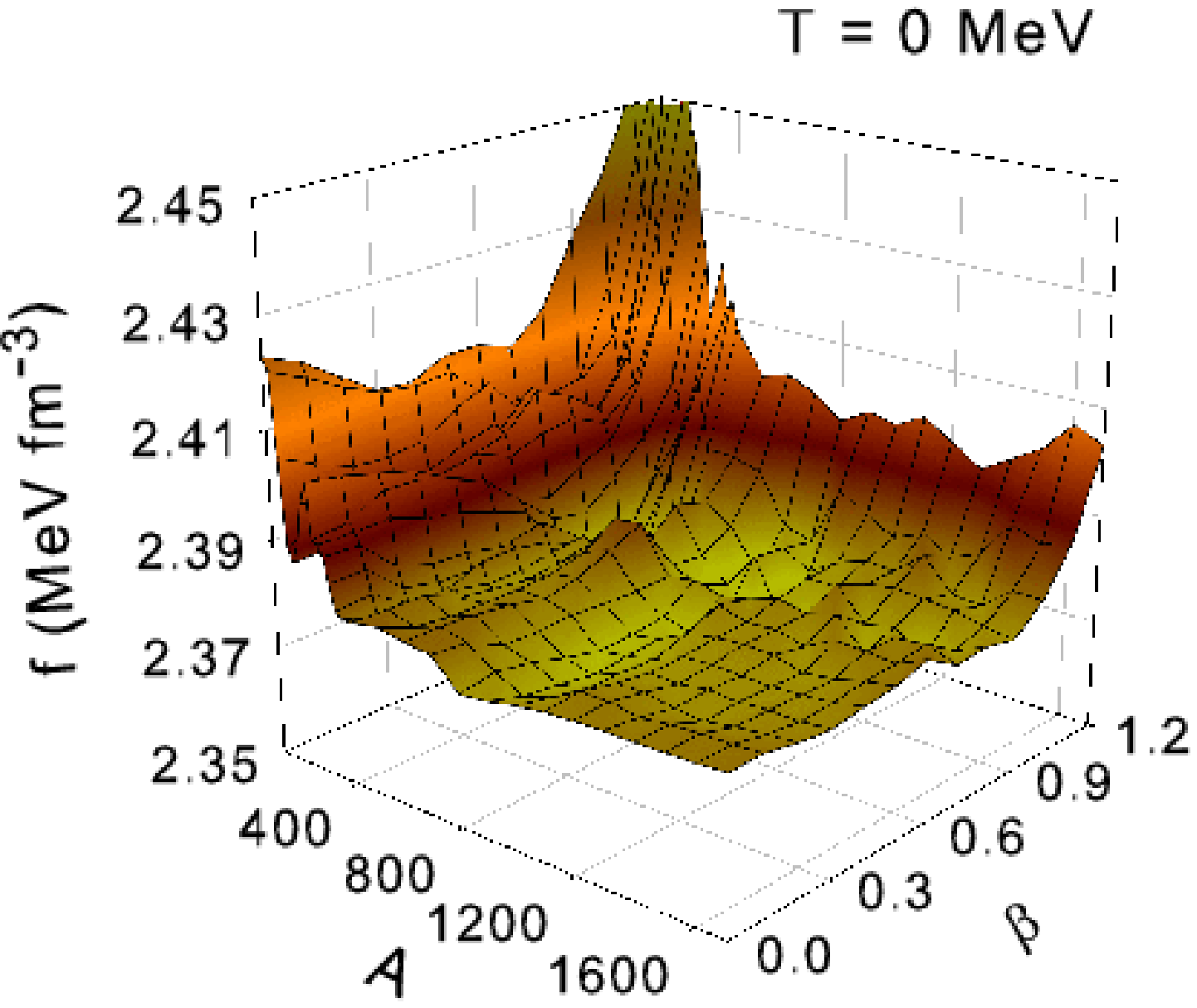,width=5.8cm}\psfig{file=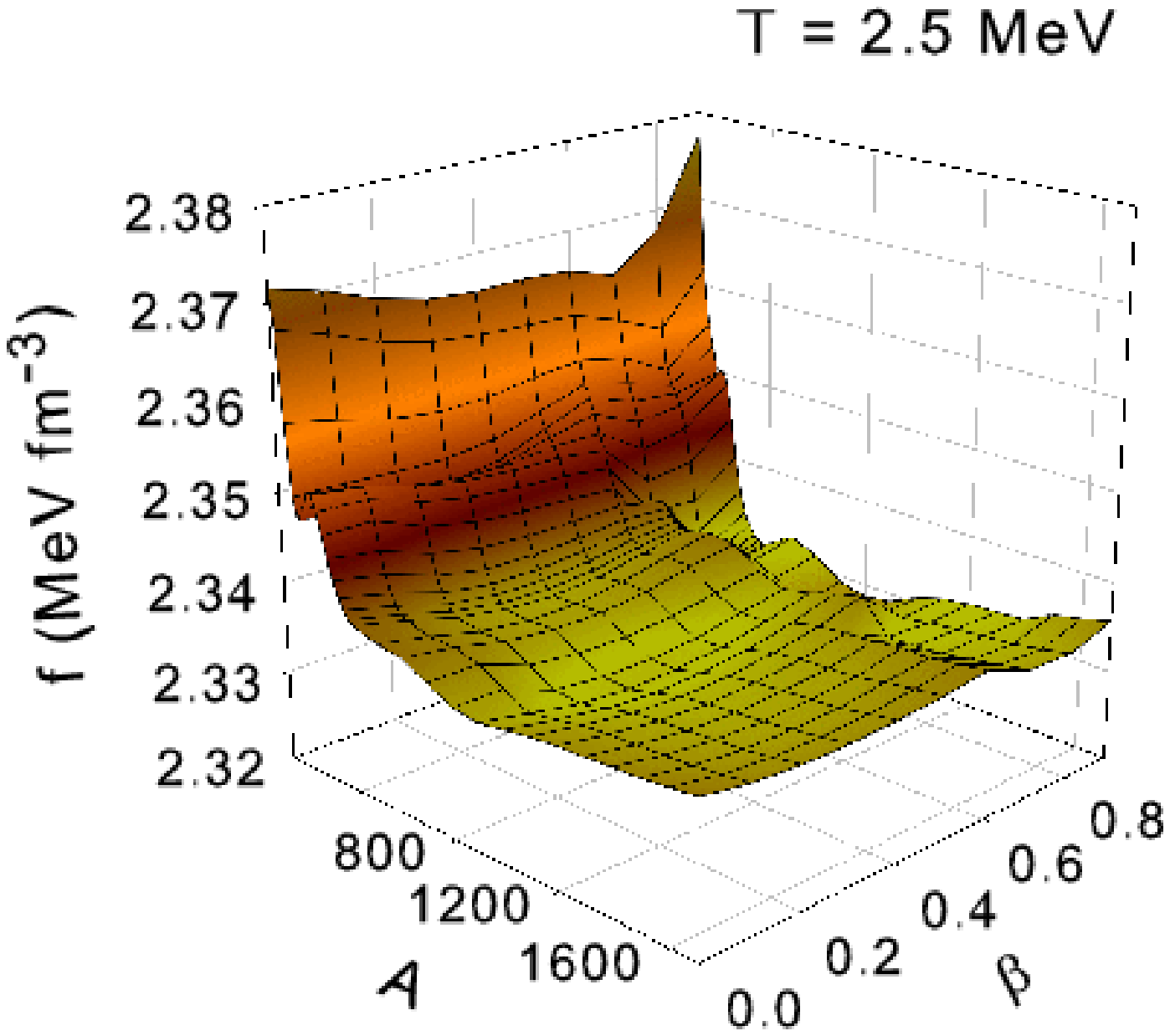,width=5.8cm}}
\centerline{\psfig{file=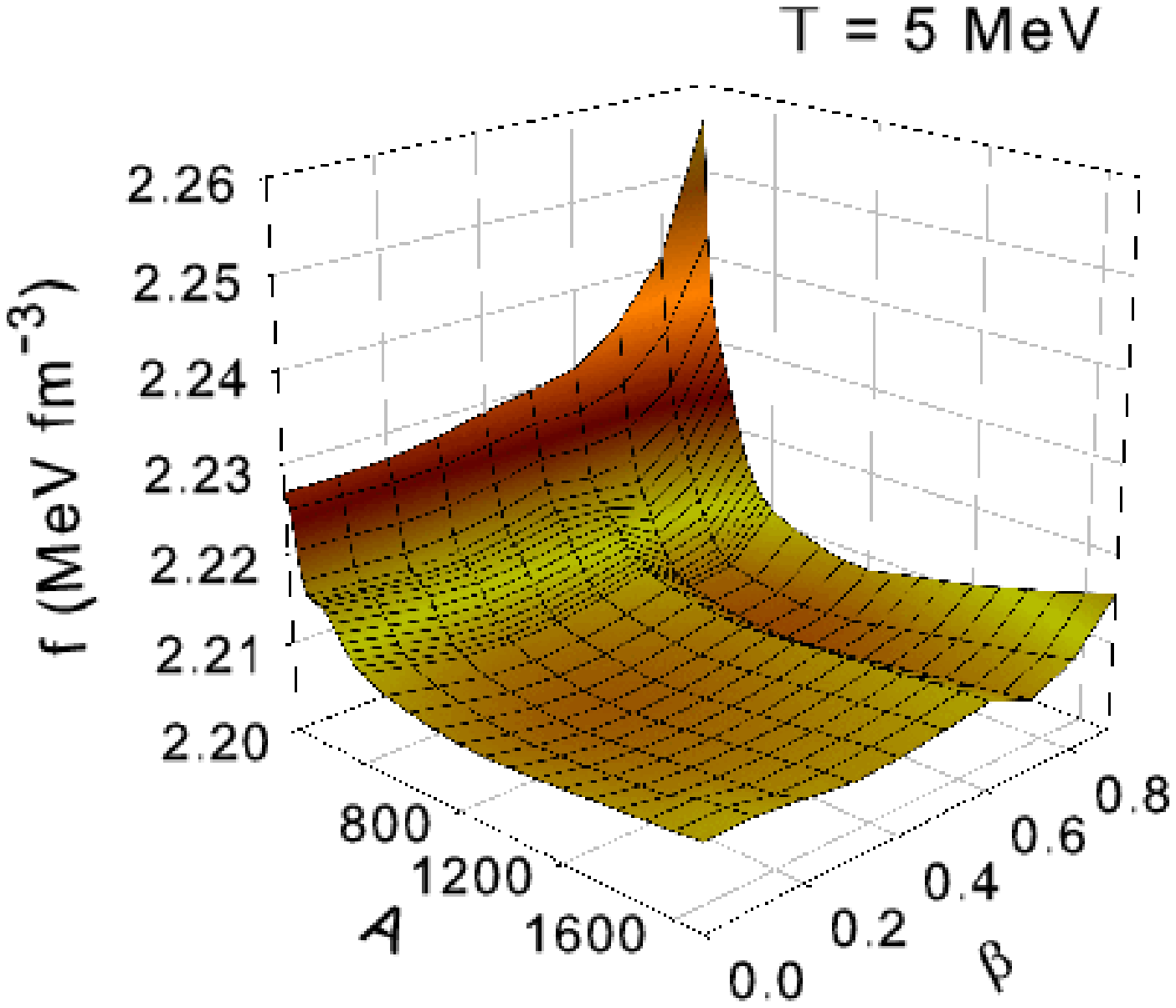,width=5.8cm}\psfig{file=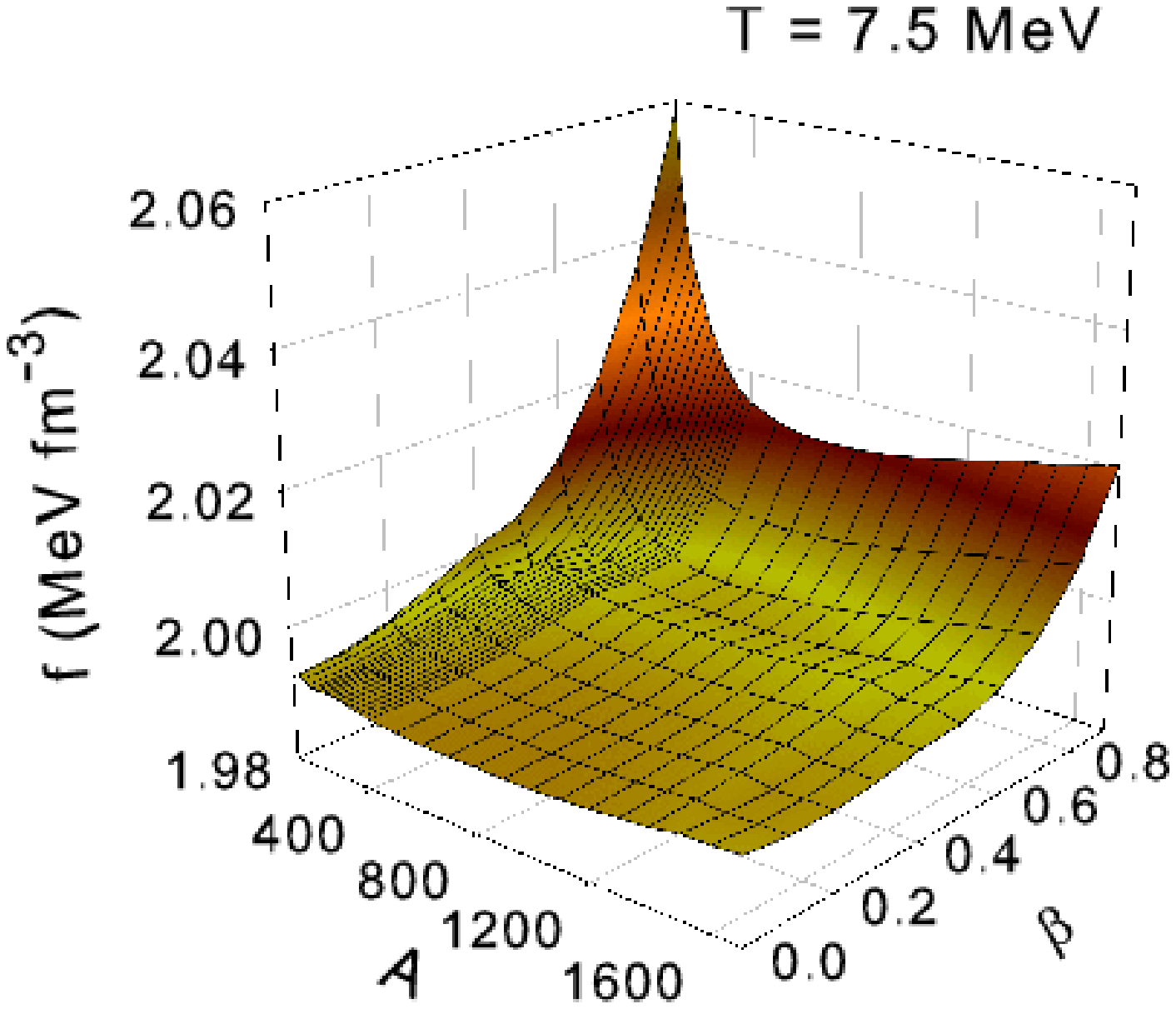,width=5.8cm}}
\caption{(Color on-line) Energy-deformation surfaces at n$_{\rm b}$ = 0.08 fm$^{\rm -3}$, $\gamma$ = $60^{\rm o}$ and $T = 0$ MeV (top left), $T = 2.5$ MeV (top right), $T = 5.0$ MeV (bottom left) and $T = 7.5$ MeV (bottom right) as calculated with the Skyrme parameterization SkM$^*$.} \hspace{1pc} \label{Fig:6}
\end{figure}

Fig.~\ref{Fig:6} displays the energy-deformation surfaces obtained at the density values n$_{\rm b}$ = 0.08 fm$^{\rm -3}$ and $\gamma$ = 60$^{\rm o}$ for four values of temperature $T$: 0, 2.5, 5.0 and 7.5 MeV. It can be seen that the energy-deformation surfaces retain the principal features, discussed in the previous section, along the curves of constant deformation. The physical shell effects associated with bound nucleons are still visible at lower temperatures. Deformation of nuclear structure breaks some of the degeneracy of the single particle energies within particular shells, leading to a larger number of sub-shells; this is seen on the energy-deformation surfaces as the increasing prominence of shell effects as $\beta$ is increased. The result of the shell effects is many local minima in the surfaces at zero or low temperature, of order of 5 keV fm$^{\rm -3}$ deep. The variation in shell energy with respect to $\beta$ at constant $A$ is expected to be caused by the varying shell structure of the bound nucleons. Magierski and Heenen \cite{magie02} suggest that in \emph{neutron star} matter, the shell effects of the unbound neutrons (caused by their scattering off the nuclear structure) contribute significantly to the variation of energy with deformation; however, this is not expected to be the case here because the density of unbound neutrons is an order of magnitude smaller in SN matter than in NS matter at a corresponding density.

\begin{figure}[!b]
\hspace{1pc}
\centerline{\psfig{file=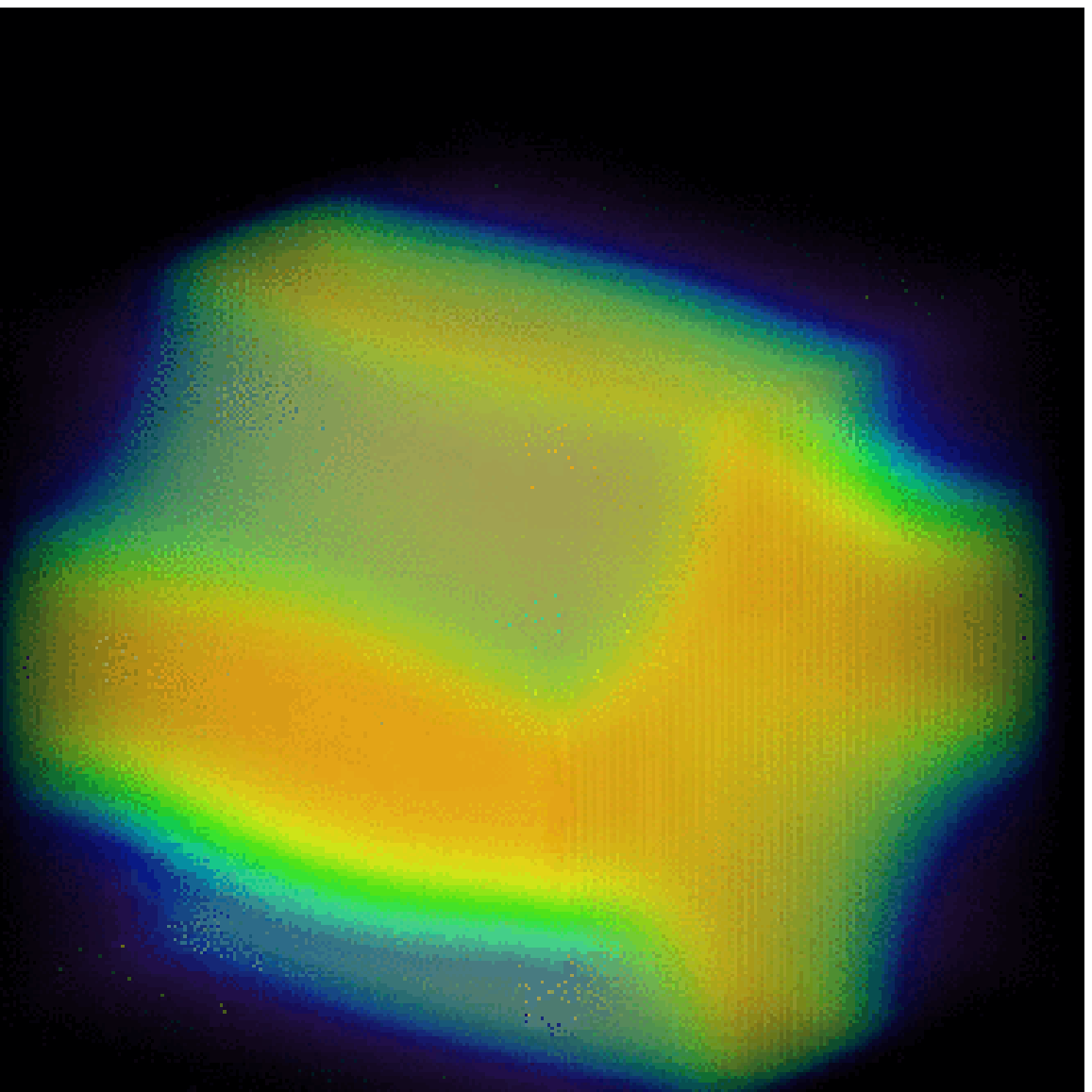,width=3.5cm}\psfig{file=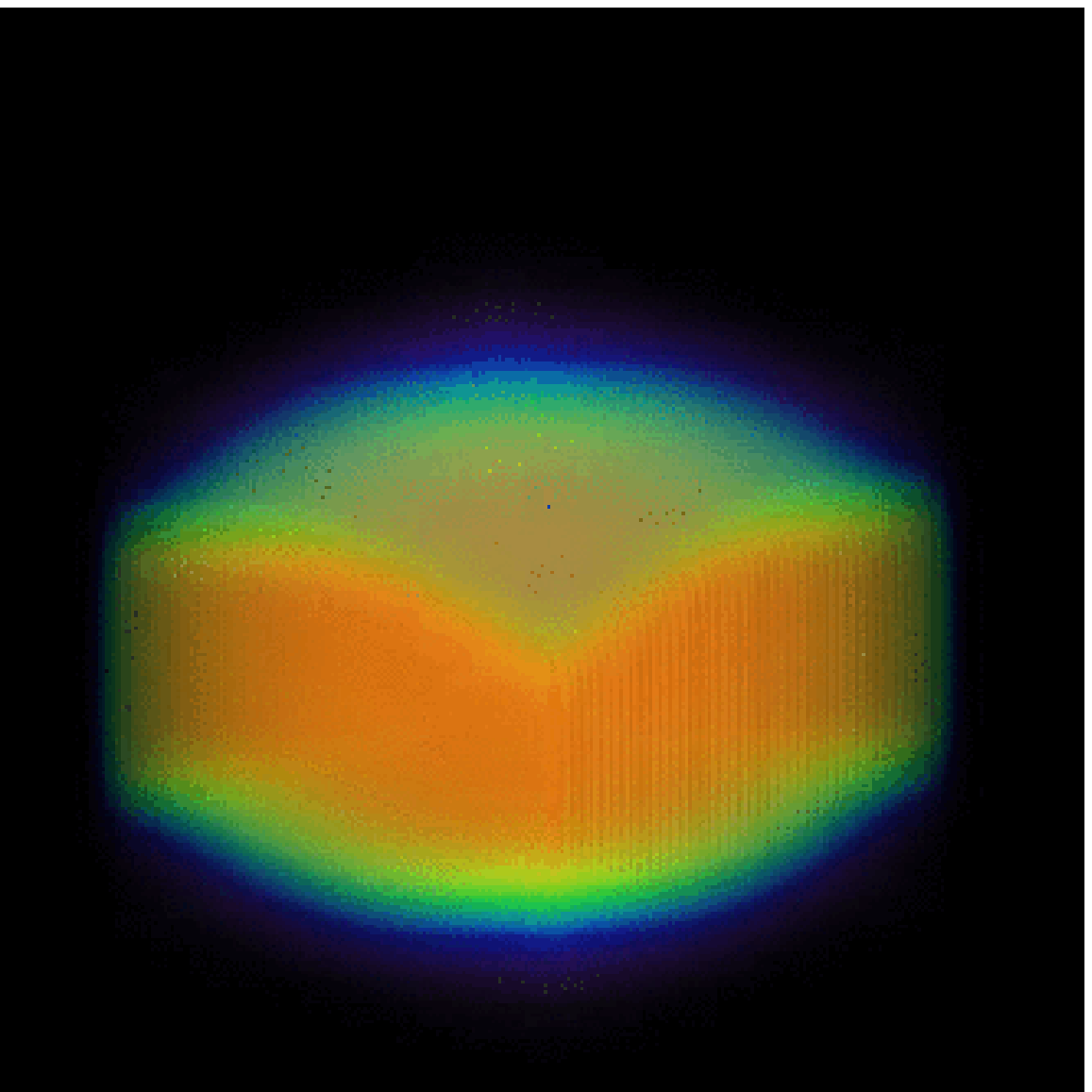,width=3.5cm}}
\caption{(Color) 3D renderings of neutron density distributions found at the broad minima on the $n_{\rm b}$ = 0.08 fm$^{\rm -3}$, $T$ = 5.0 MeV; $\gamma$=$60^{\rm o}$ energy surface. The left configuration, a cylindrical hole, is found at $\beta = 0.3, A = 800$ and the right configuration, a slab-like structure, is found at $\beta \approx 0.7, A \approx 800$. Blue indicates the lowest densities and red the highest.} \hspace{1pc} \label{Fig:7}
\end{figure}

The shell effects become washed out at higher temperature leaving behind the geometrical minima. We see that these occur not only with respect to $A$, but also with respect to $\beta$, and result in several local geometrical minima appearing on each energy-deformation surface. For example, at $T$ = 5.0 MeV, $n_{\rm b}$ = 0.08 fm$^{\rm -3}$ and $\gamma = 60^{\rm o}$ (Fig.~\ref{Fig:6}; bottom left) there are two broad minima that appear, centered at $\beta \approx 0.3, A \approx 800$ and $\beta \approx 0.7, A \approx 800$. The geometrical configurations they correspond to are displayed in Fig.~\ref{Fig:7}. The former corresponds to slab like nuclear shape (`lasagna'); the latter to a cylindrical hole nuclear shapes. They are separated by a `ridge' in the surface, running along $\beta = 0.6$ which delineates the regions on the energy surface where the two geometries occur. The free energy-density changes smoothly in the vicinity of the minima, and the nuclear geometry locally remains the same.

\begin{figure}[!t]
\hspace{1pc}
\centerline{\psfig{file=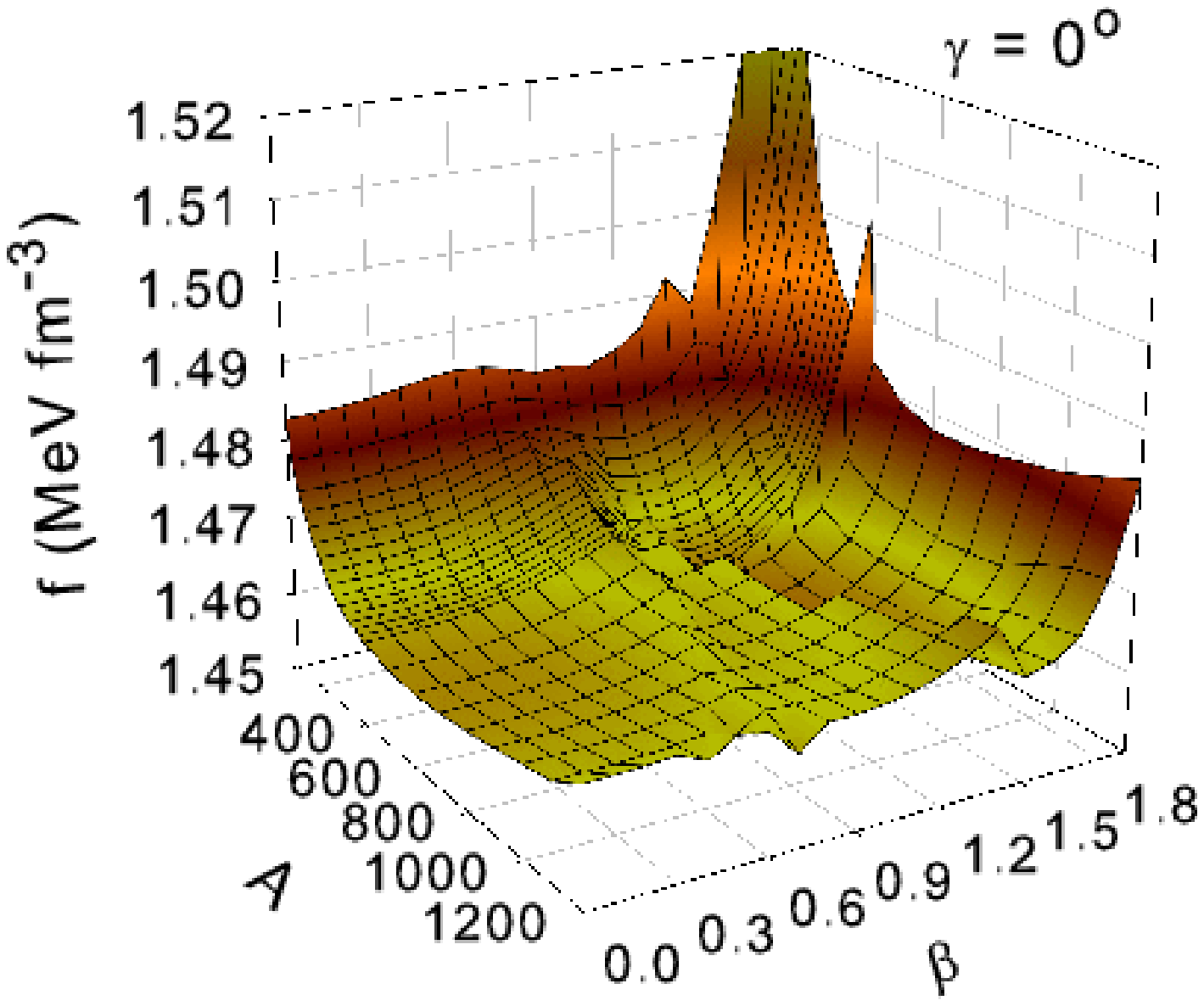,width=5cm}\psfig{file=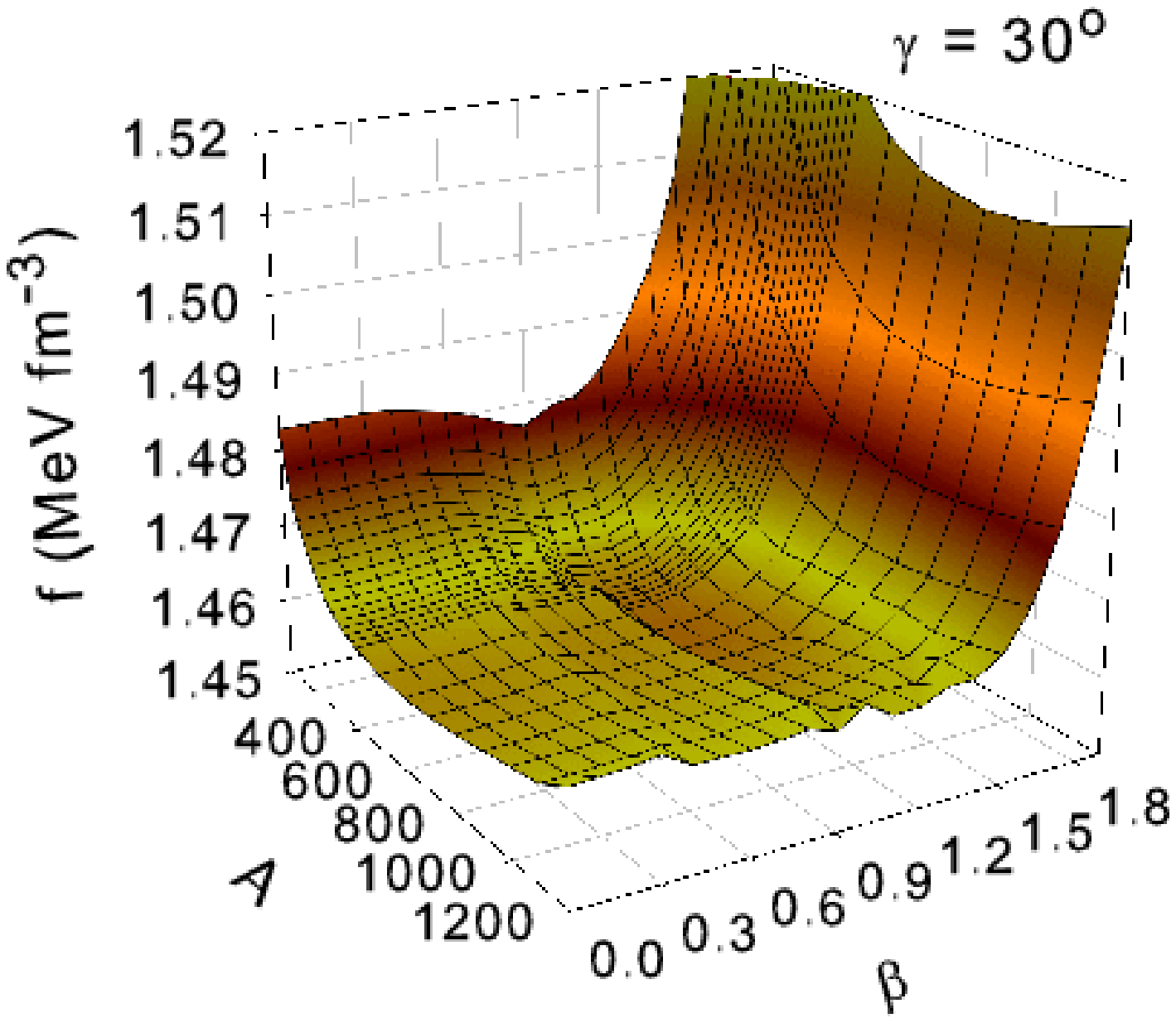,width=5cm}\psfig{file=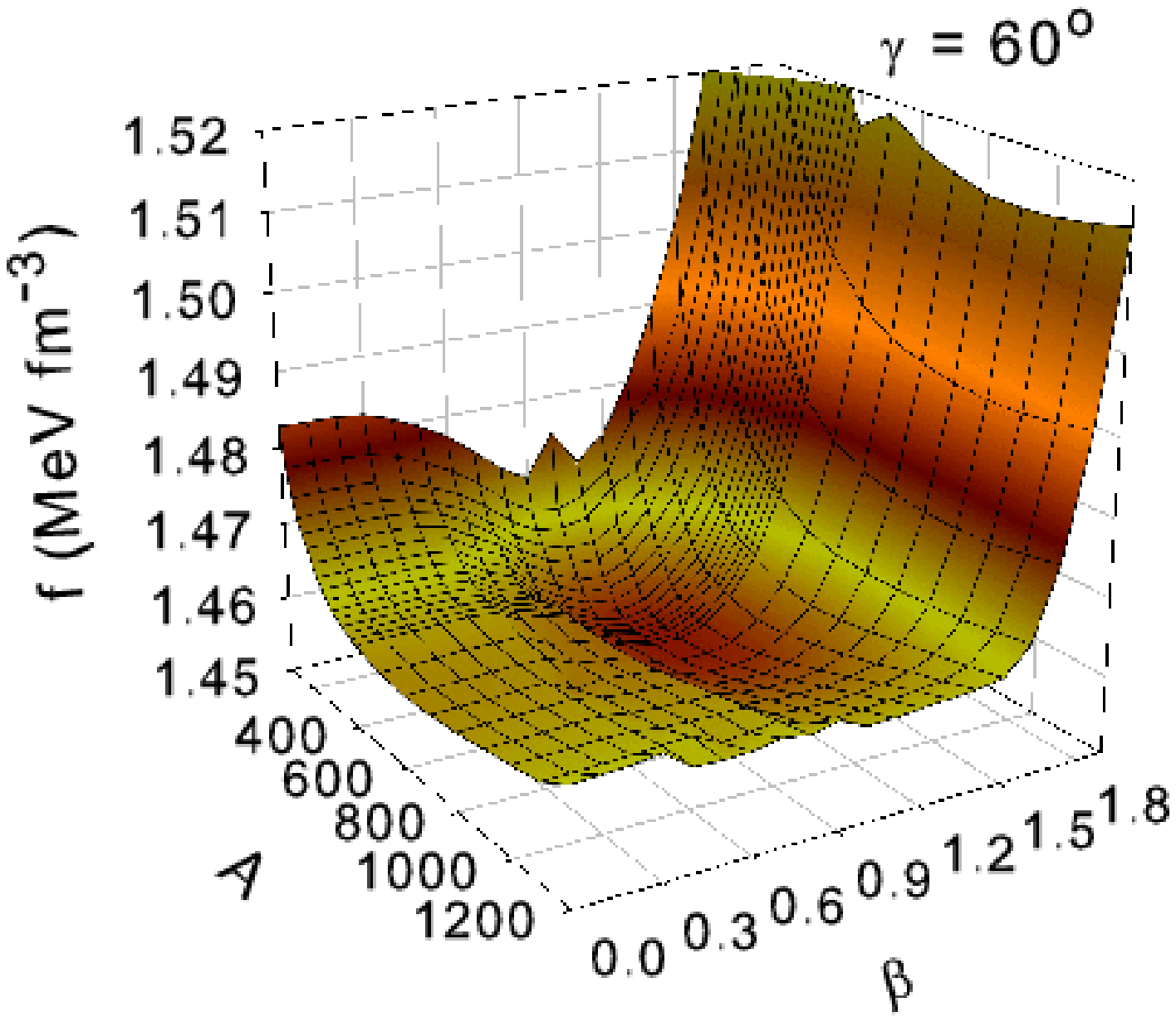,width=5cm}}
\caption{(Color on-line) Energy-deformation surfaces at $n_{\rm b}$ = 0.06 fm$^{\rm -3}$, $T$ = 5.0 MeV; $\gamma$=$0^{\rm o}$ (left), $30^{\rm o}$ (middle) and $60^{\rm o}$ (right) as calculated with the Skyrme parameterization SkM$^*$.} \hspace{1pc} \label{Fig:8}
\end{figure}

Fig.~\ref{Fig:8} shows the energy-deformation surfaces obtained at the density values $n_{\rm b}$ = 0.06 fm$^{\rm -3}$ and temperature $T$ = 5.0 MeV for three different values of the deformation direction $\gamma$: 0$^{\rm o}$, 30$^{\rm o}$ and 60$^{\rm o}$. The minimum with respect to $A$ occurs at 600-800 and is fairly constant with respect to $\beta$. As $\gamma$ changes, so do the geometries obtained as $\beta$ increases. A deep minimum appears on the $\gamma = 60^{\rm o}$ energy surface at $A \approx 600$, $\beta = 1.1$, and indeed this corresponds to the absolute minimum found at this density and temperature. It is important to note, however, that the energy separations between local minima is relatively small (of order keV) and one must expect multiple configurations to exist at a given density and temperature.

\begin{figure}[!b]
\hspace{1pc}
\centerline{\psfig{file=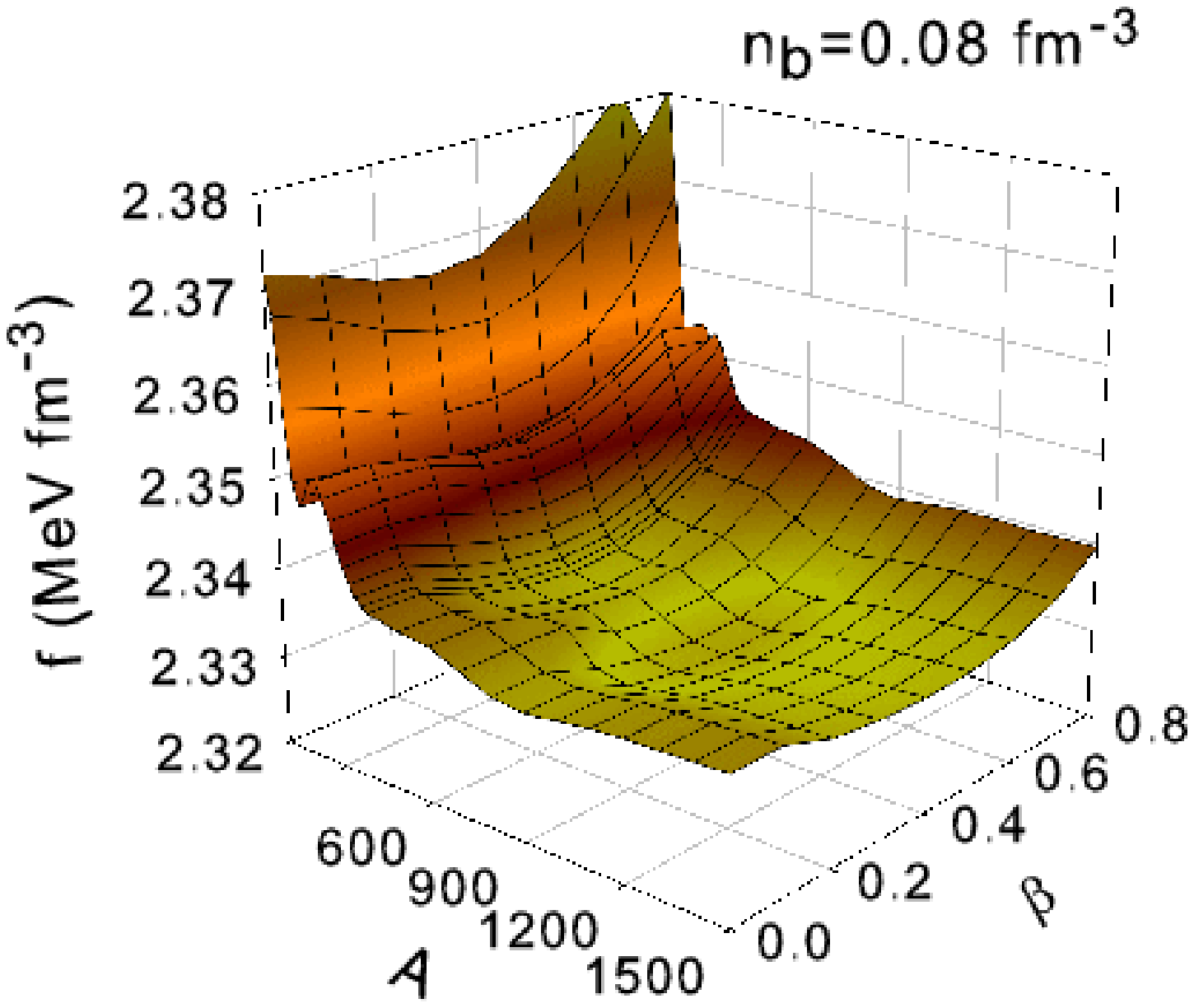,width=5cm}\psfig{file=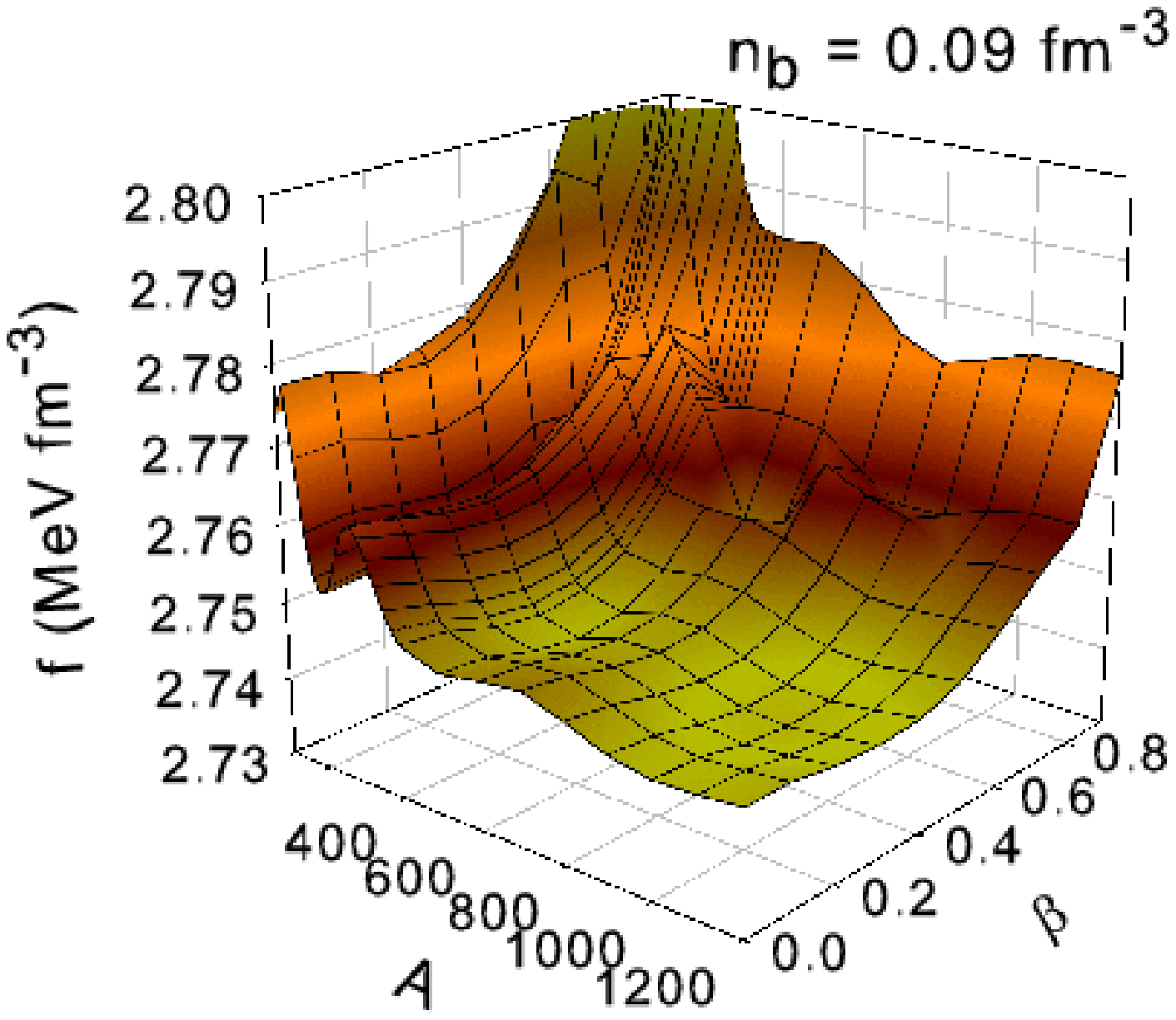,width=5cm}\psfig{file=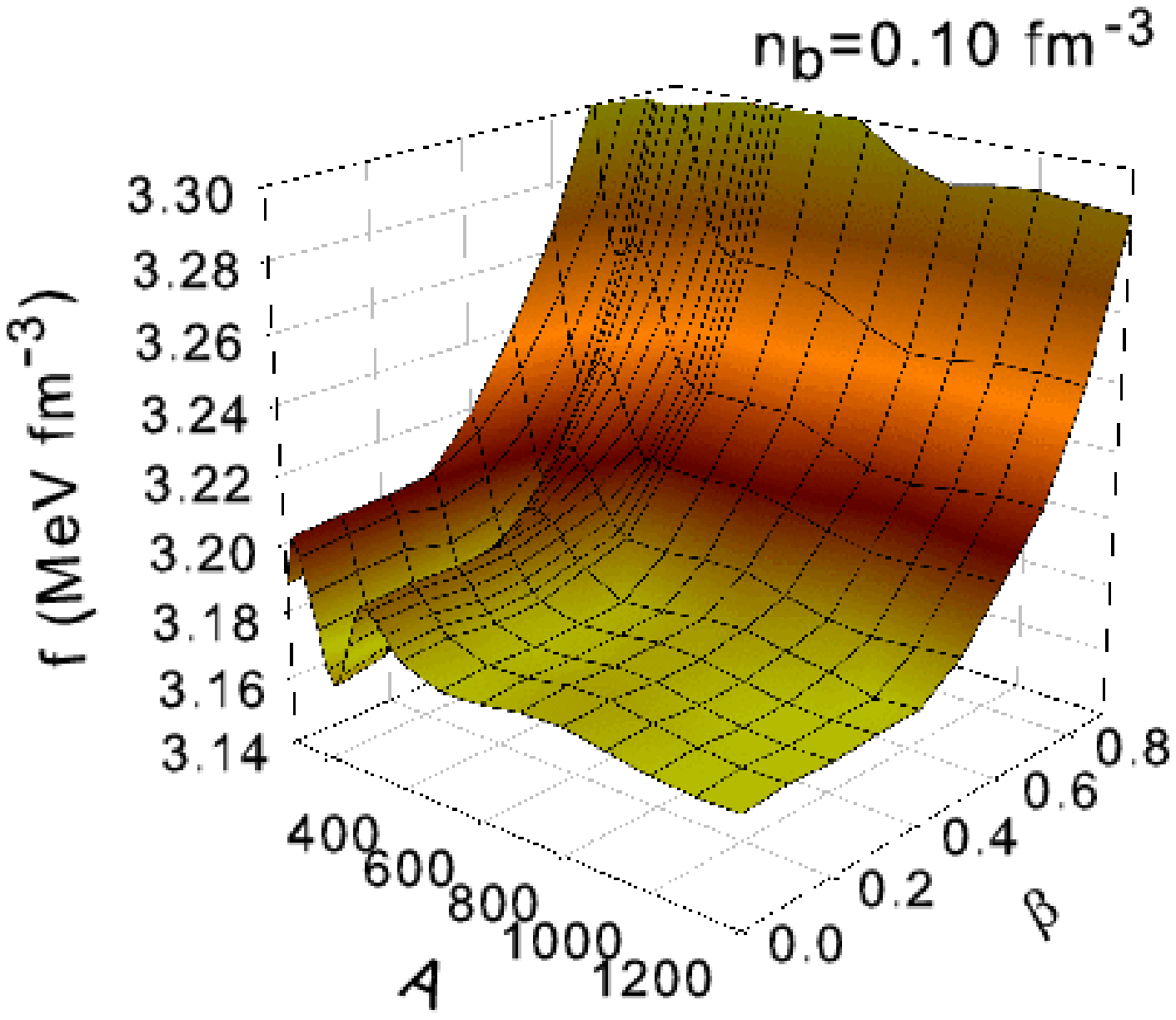,width=5cm}}
\caption{(Color on-line) Energy-deformation surfaces for $T$ = 2.5 MeV, $\gamma$ = 0$^{\rm o}$ and n$_{\rm b}$ = 0.08 fm$^{\rm -3}$ (left), n$_{\rm b}$ = 0.09 fm$^{\rm -3}$ (middle) and $n_{\rm b}$ = 0.10 fm$^{\rm -3}$ (right).} \hspace{1pc} \label{Fig:9}
\end{figure}

Fig.~\ref{Fig:9} shows the energy-deformation surfaces obtained at the densities n$_{\rm _b}$ = 0.08 - 0.10 fm$^{\rm -3}$ for $\gamma$ = 0$^{\rm o}$ and temperature $T = $ 2.5 MeV. As the density increases the nature of the shell effects is seen to change, becoming dominated by a spurious component the highest density and low deformations. The free energy-density rises increasingly sharply for $\beta \ge$ 0.4 as density increases. This is a clear indication that the favored ground state is becoming closer to uniform matter; it is costing more energy to deform the matter into structure. It is also interesting to note that the spurious shell effects disappear at large deformation, as the structure becomes far from uniform again.

\subsection{The EoS and the Transition to Uniform Matter}

By taking the minimum free energy-density configurations from each baryon number density, and temperature, the EoS of non-uniform matter can be constructed. We have performed this minimization over $A$, $\beta$ and three values of $\gamma$ (0$^{\rm o}$, 30$^{\rm o}$, 60$^{\rm o}$) at seven different values of density: $n_{\rm b}$ = 0.04, 0.06, 0.08, 0.09, 0.10, 0.11, 0.12 fm$^{-3}$ and three different values of temperature $T$ = 0, 2.5 and 5 MeV. One should bear in mind that spurious shell effects might alter the position of the minima in phase space at densities and temperatures close the transition to uniform matter.

In Tables~\ref{Tab:3} to~\ref{Tab:4} the variation of relevant EoS quantities with density and temperature are given. Values of $A$, $\beta$ and $\gamma$ corresponding to the minima are tabulated, together with the corresponding values of free energy-density $f$, entropy density $s$, pressure $p$, proton, neutron and electron chemical potentials $\mu_{\rm p}$, $\mu_{\rm n}$, $\mu_{\rm e}$, and the quantity $\delta = r_{\rm RMS} / r_{\rm C}$, where $r_{\rm RMS}$ is the nucleon root mean square radius and $r_{\rm C}$ is equal to half the width of the cell. $\delta$ is a measure of the fraction of the cell occupied by the nuclear structure; for uniform matter, $\delta = 1$.

The energy minima appear at values of $A$ that are generally smaller than those found by BV \cite{bonche81} (e.g. at $T$ = 5 MeV, $n_{\rm b}$ = 0.06 fm$^{-3}$, a minimum at $A = 600$ is found as opposed to $A$ = 900 for BV). This is a result of allowing deformed nuclear structure; similar values of $A$ to BV are obtained if one only admits zero deformation configurations (it was observed in Sec.~\ref{sec3ss1} that at $n_{\rm b}$ = 0.06 fm$^{-3}$, $T = 0 - 2.5$ MeV, $(\beta, \gamma)$ = (0.0,0$^{\rm o}$), the same value of $A$ as BV for the minimum energy was obtained).

\begin{figure}[!t]
\hspace{1pc}
\centerline{\psfig{file=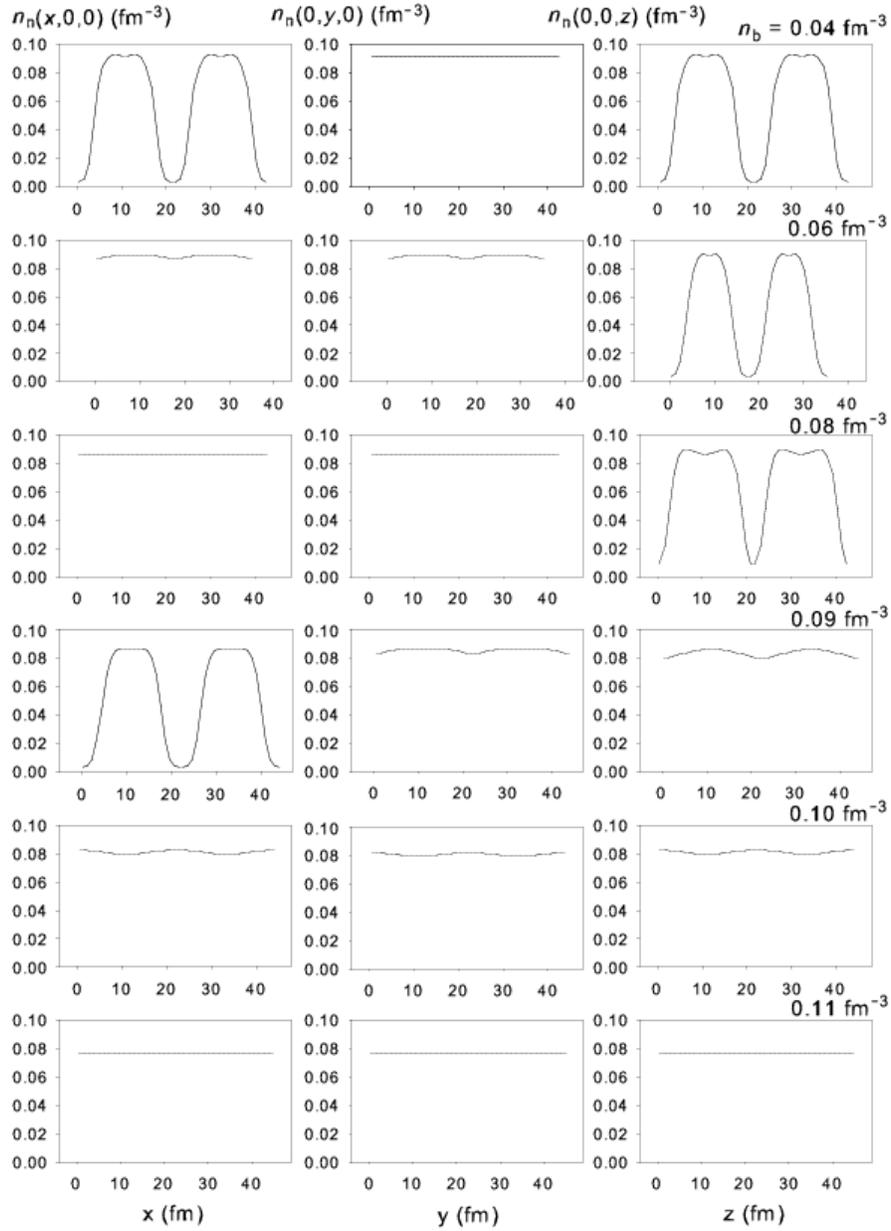,width=12cm}}
\caption{Neutron density profiles $n_{\rm n} (x,y,z)$ of minimum energy configurations along a line joining the centers of adjacent cells in three perpendicular directions $x, y$ and $z$ for the six densities and at a temperature of 2.5 MeV.} \hspace{1pc} \label{Fig:10}
\end{figure}

In Fig.~\ref{Fig:10} the neutron density profiles are shown along a line connecting two adjacent unit cells in three perpendicular directions for the minimum energy configurations at all seven densities and $T$ = 2.5 MeV. The axes are labeled $x, y$ and $z$, but one should note that the choice of axis labeling is arbitrary in relation to the macrophysical structure of the matter. In Fig.~\ref{Fig:11} 3D renderings of those neutron density distributions in one unit cell are displayed. It can be seen that the transition proceeds to uniform matter proceeds along the recognised ordering of the pasta shapes: cylindrical, planar, cylindrical hole and spherical hole. We finally achieve a self-consistent transition to uniform matter from increasing the density at around $n_{\rm b}$ = 0.10 - 0.11 fm$^{-3}$ at this temperature.

In Fig.~\ref{Fig:12} 3D renderings of the neutron density distribution of energy-minimum configurations taken at $n_{\rm b}$ = 0.10 fm$^{-3}$, for $T$ = 0, 2.5, 5, and 7.5 MeV are displayed. Up to $T$ = 5 MeV, the configuration is bubble-like; at $T$ = 7.5 MeV a self-consistent transition to uniform matter has been reached by increasing the temperature.

Despite the fact that the canonical pasta shapes were obtained at the absolute minima in energy at the selected values of density, it should be noted that a far richer variety of nuclear shapes were obtained here at local minima close to the absolute ground state. It is expected that, as calculations are performed over a greater number of densities in between those already studied, the transition between the canonical pasta shapes will be mediated by a number of different nuclear geometries. In addition, it is likely that at a single given density and temperature, one will find many pasta shapes co-existing. The implication is that at a macrophysical level, the bulk properties of matter will be given by an averaging over the properties of those individual pasta phases, and that the thermodynamical properties of matter will change smoothly with density and temperature through the pasta regime until the phase transition to uniform matter is reached.

\begin{figure}[!t]
\centerline{\psfig{file=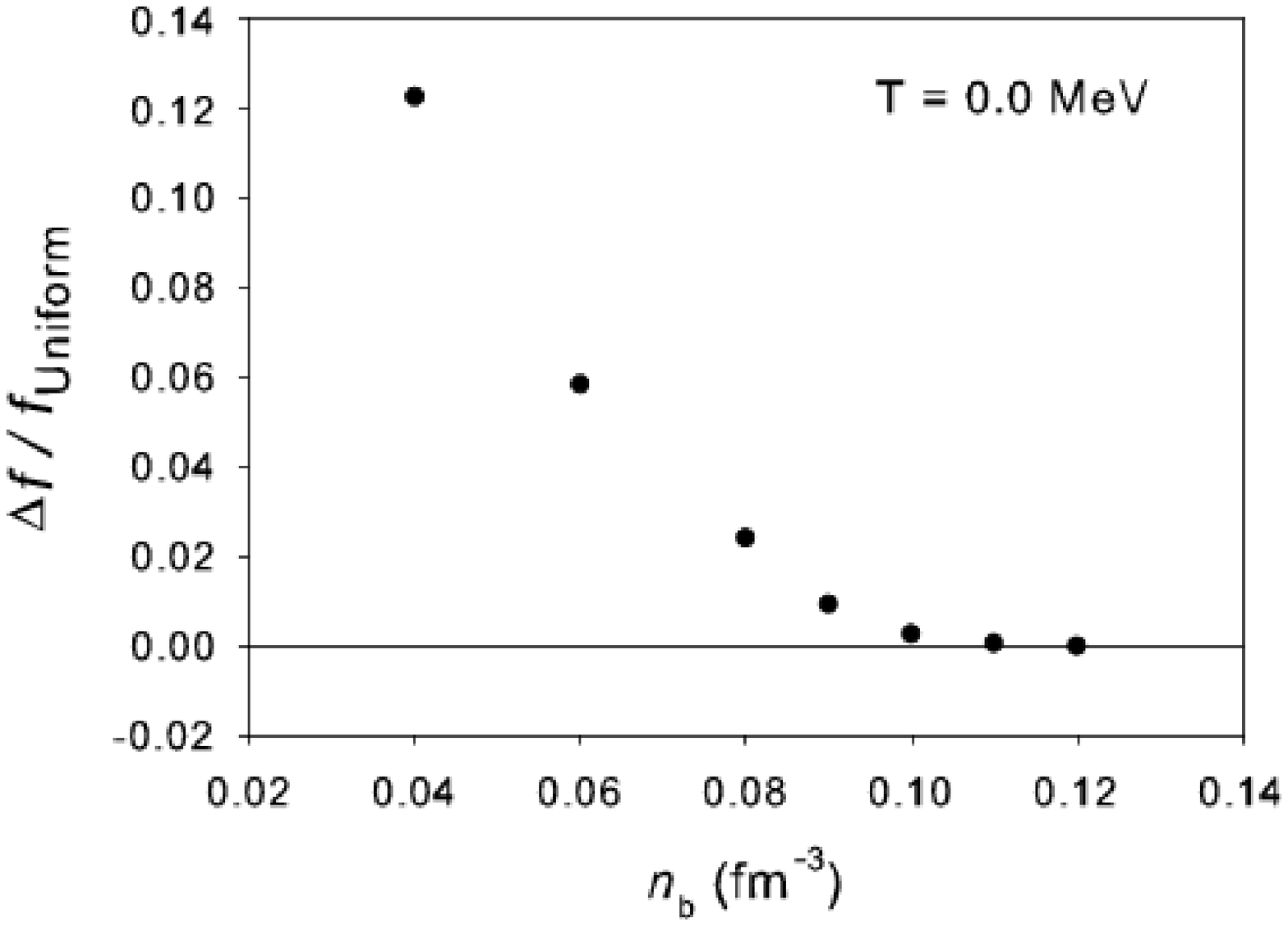,width = 9.0cm}\psfig{file=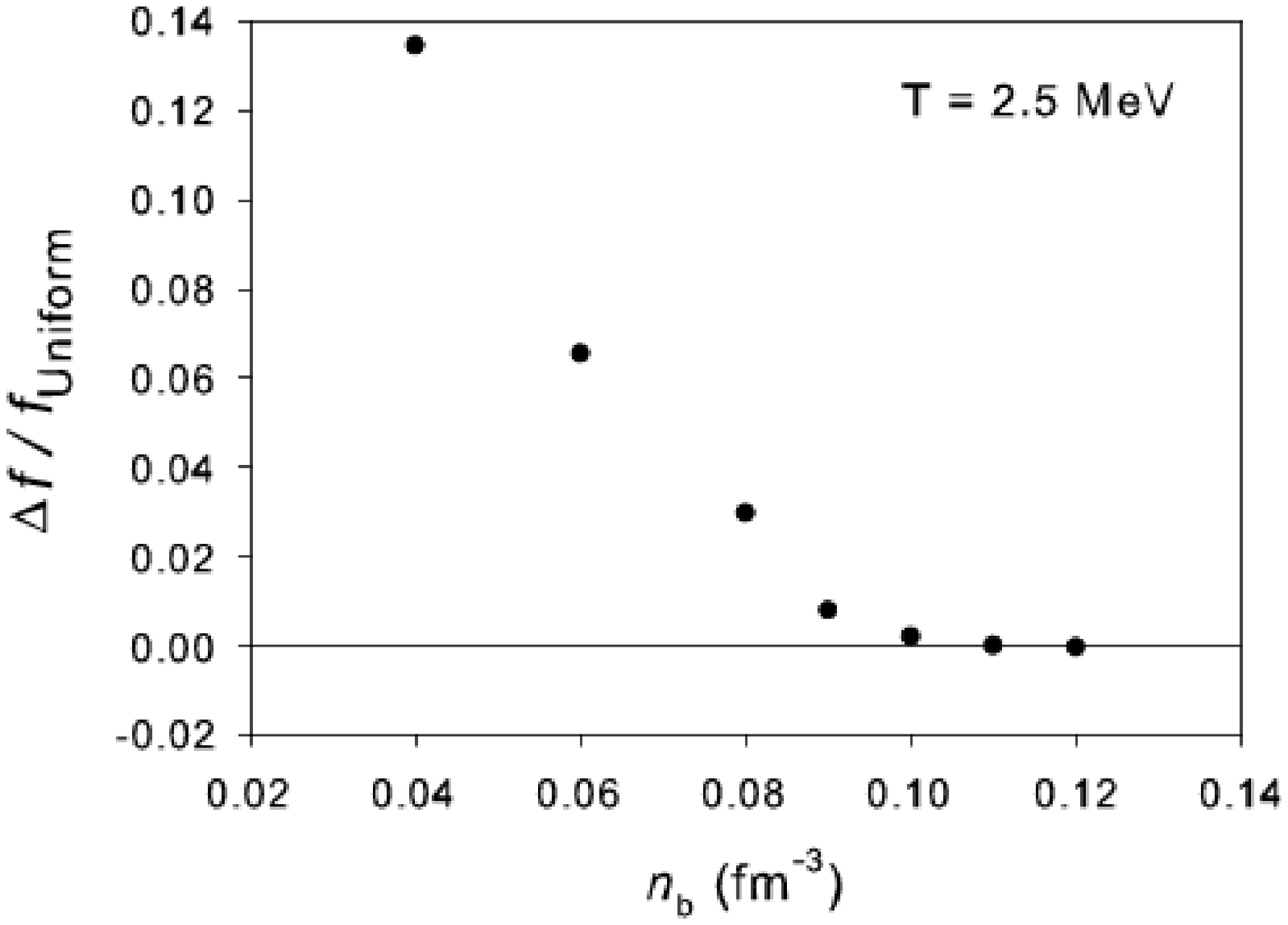,width = 9.0cm}}
\centerline{\psfig{file=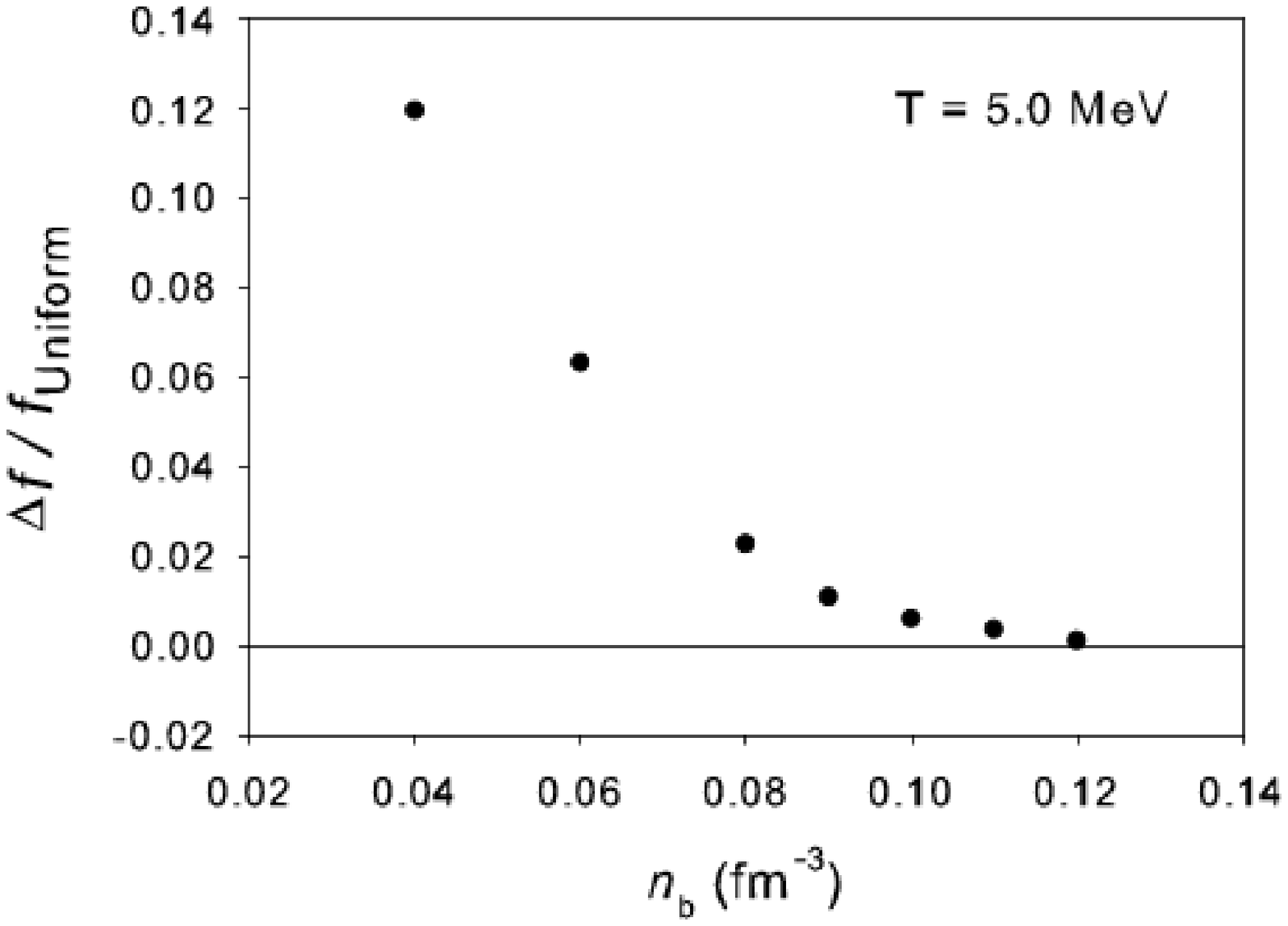,width = 9.0cm}}
\caption{Difference between the free energy-density of uniform matter and non-uniform matter $\Delta f$ divided by the free energy-density of non-uniform matter $f_{\rm Uniform}$ as density increases for $T$ = 0 MeV (top left), 2.5 MeV (top right) and 5 MeV (bottom).}
\label{Fig:13}
\end{figure}

\begin{figure}[!t]
\centerline{\psfig{file=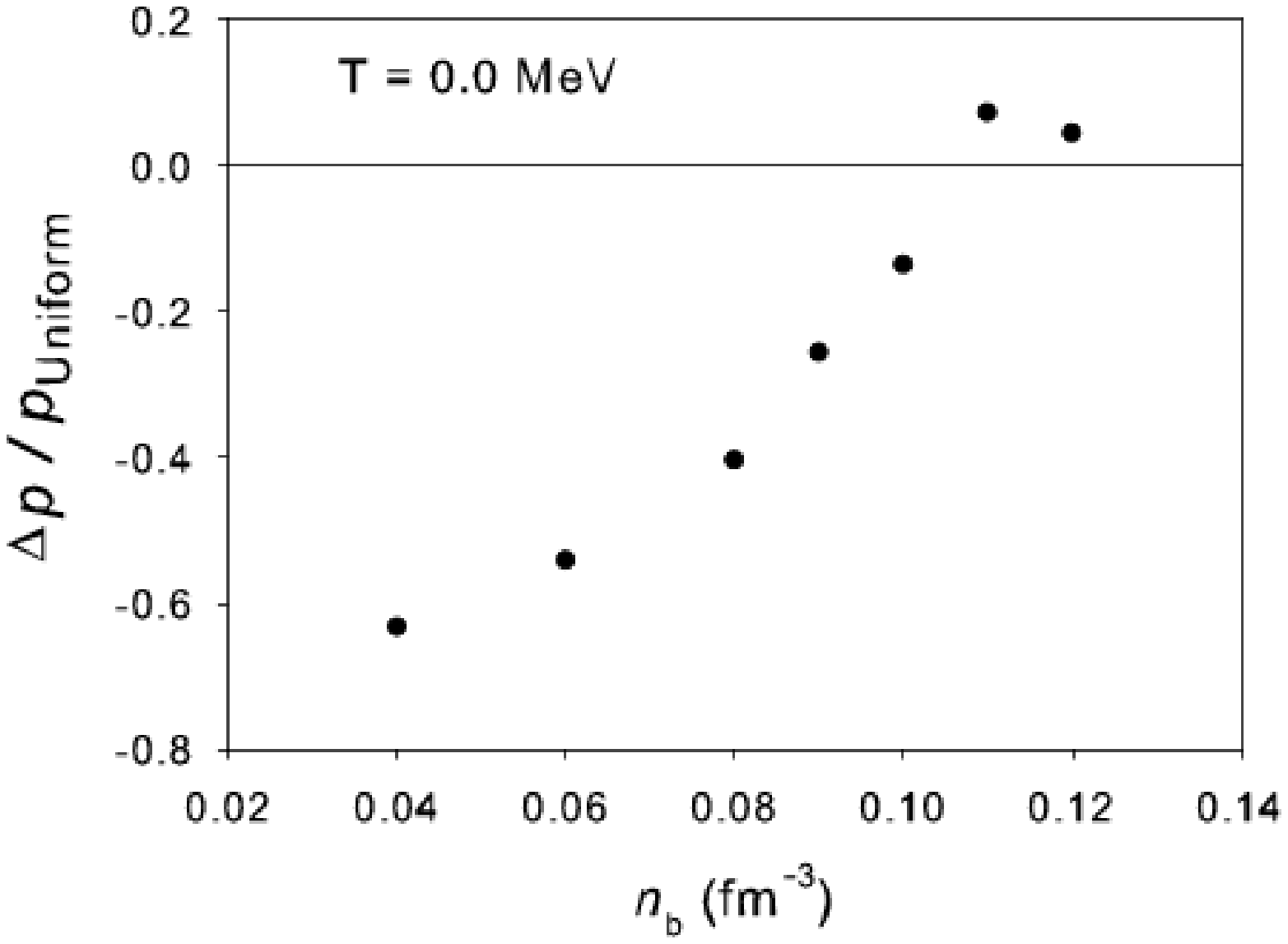,width = 9.0cm}\psfig{file=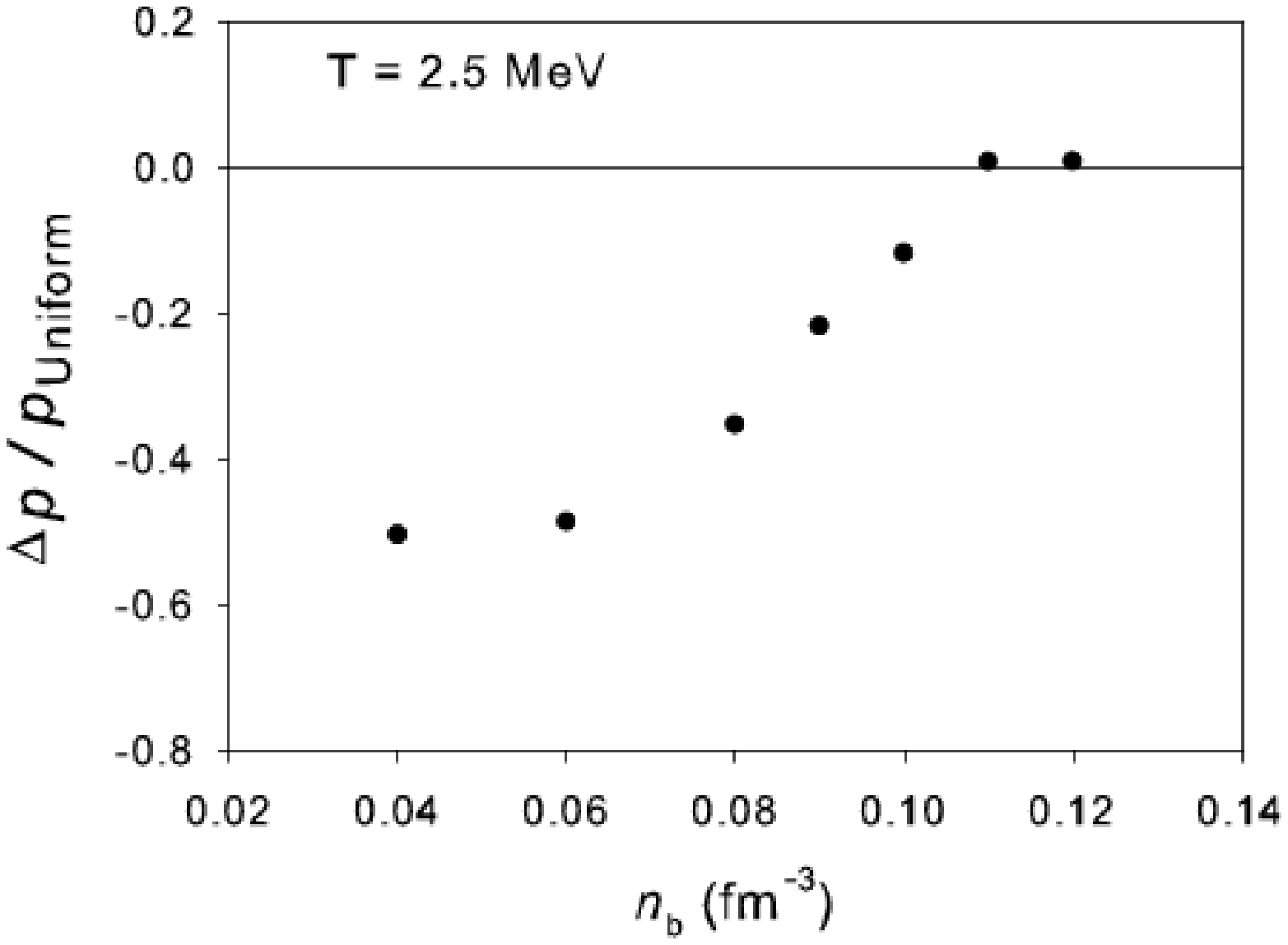,width = 9.0cm}}
\centerline{\psfig{file=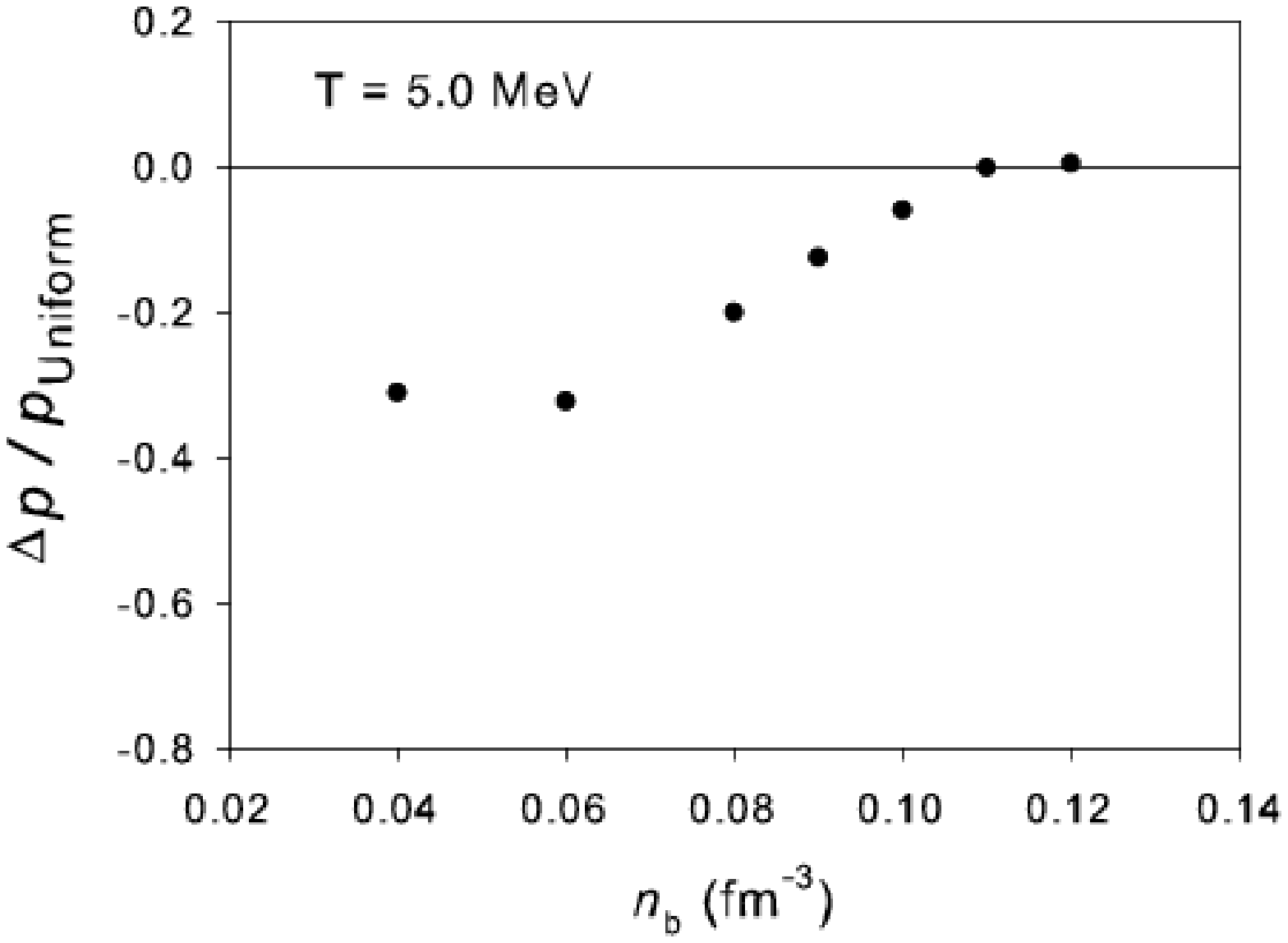,width = 9.0cm}}
\caption{Difference between the pressure of uniform matter and non-uniform matter $\Delta p$ divided by the pressure of non-uniform matter $p_{\rm Uniform}$ as density increases for $T$ = 0 MeV (top left), 2.5 MeV (top right) and 5 MeV (bottom).}
\label{Fig:14}
\end{figure}

\begin{figure}[!b]
\centerline{\psfig{file=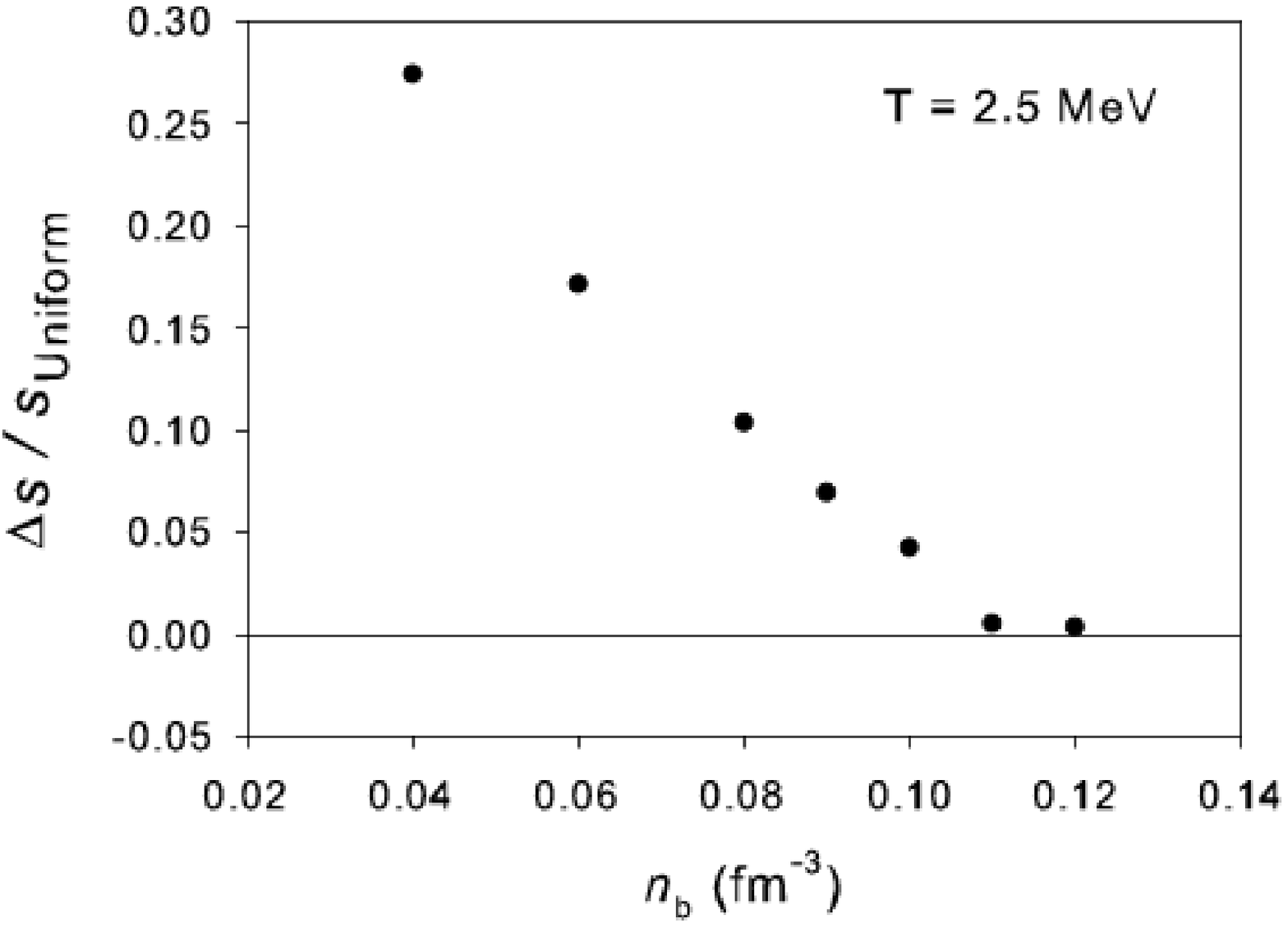,width = 9.0cm}\psfig{file=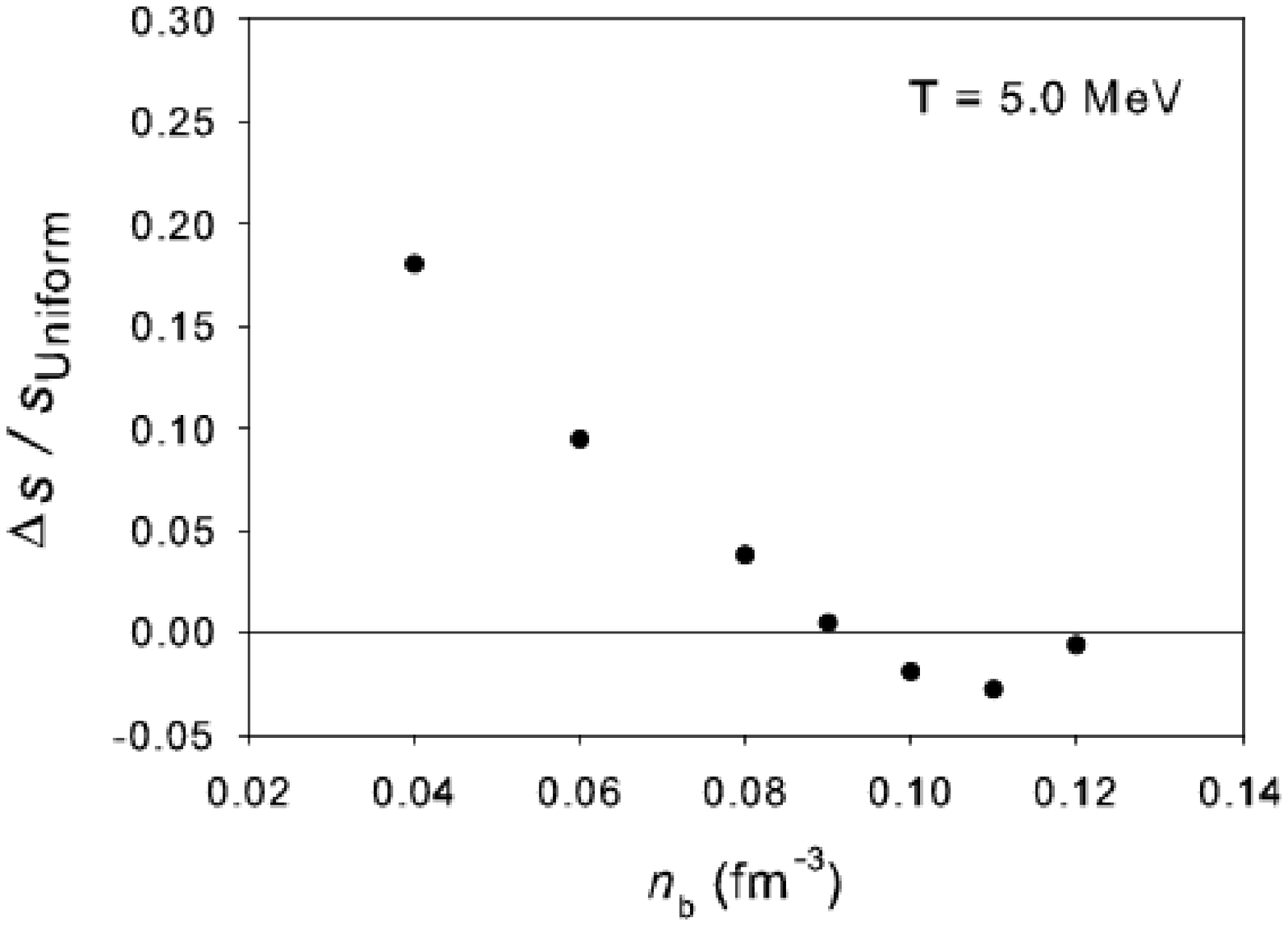,width = 9.0cm}}
\caption{Difference between the entropy density of uniform matter and non-uniform matter $\Delta s$ divided by the entropy density of non-uniform matter $s_{\rm Uniform}$ as density increases for $T$ = 2.5 MeV (left) and 5 MeV (right).}
\label{Fig:15}
\end{figure}

For the present model to produce the transition to uniform matter consistently, the various thermodynamic quantities calculated should tend in some way towards those calculate in the uniform matter where the Skyrme energy-density is given analytically. In order to check this, the difference in free energy-density, entropy density and pressure between the inhomogeneous matter and the uniform matter Skyrme-HF model is plotted in Figs.~\ref{Fig:13} to~\ref{Fig:15} for the densities already considered and temperatures of 0 MeV, 2.5 MeV and 5 MeV. The difference in free energy-density $\Delta f / f_{\rm Uniform} = (f_{\rm Uniform} - f_{\rm Non-Uniform})/f_{\rm Uniform}$ is displayed in Fig.~\ref{Fig:13} at the three temperatures and is seen to vanish relatively smoothly as the transition density $n_{\rm b} \approx 0.10 - 0.11$ fm$^{-3}$ is approached. The equivalent pressure difference $\Delta p / p_{\rm Uniform} = (p_{\rm Uniform} - p_{\rm Non-Uniform})/p_{\rm Uniform}$ and entropy difference $\Delta f / f_{\rm Uniform} = (s_{\rm Uniform} - s_{\rm Non-Uniform})/s_{\rm Uniform}$, shown in Fig.~\ref{Fig:14} and Fig.~\ref{Fig:15} (the latter only at the two non-zero temperatures) exhibit discontinuities at the transition density in a way that suggests the occurrence of a phase transition, as expected. More detailed calculations around the transition density are required to determine the exact nature of the transition. Errors incurred from the spurious shell energy will have an effect on the details of the results just below the transition density and may obscure the nature of the phase transition obtained (i.e. first order transition (discontinuity in pressure) or second order (discontinuity in the gradient of the pressure)). A more detailed investigation of the phase transition region will be presented in a subsequent work. Physically, the transition is expected to be first order, as in a liquid-gas transition. However, it can be stated with some confidence that the 3D Hartree-Fock model self-consistently predicts a first or second order phase transition from non-uniform nuclear matter to uniform nuclear matter.

\subsection{Dependence on Nuclear Force}

\begin{figure}[!t]
\hspace{1pc}
\centerline{\psfig{file=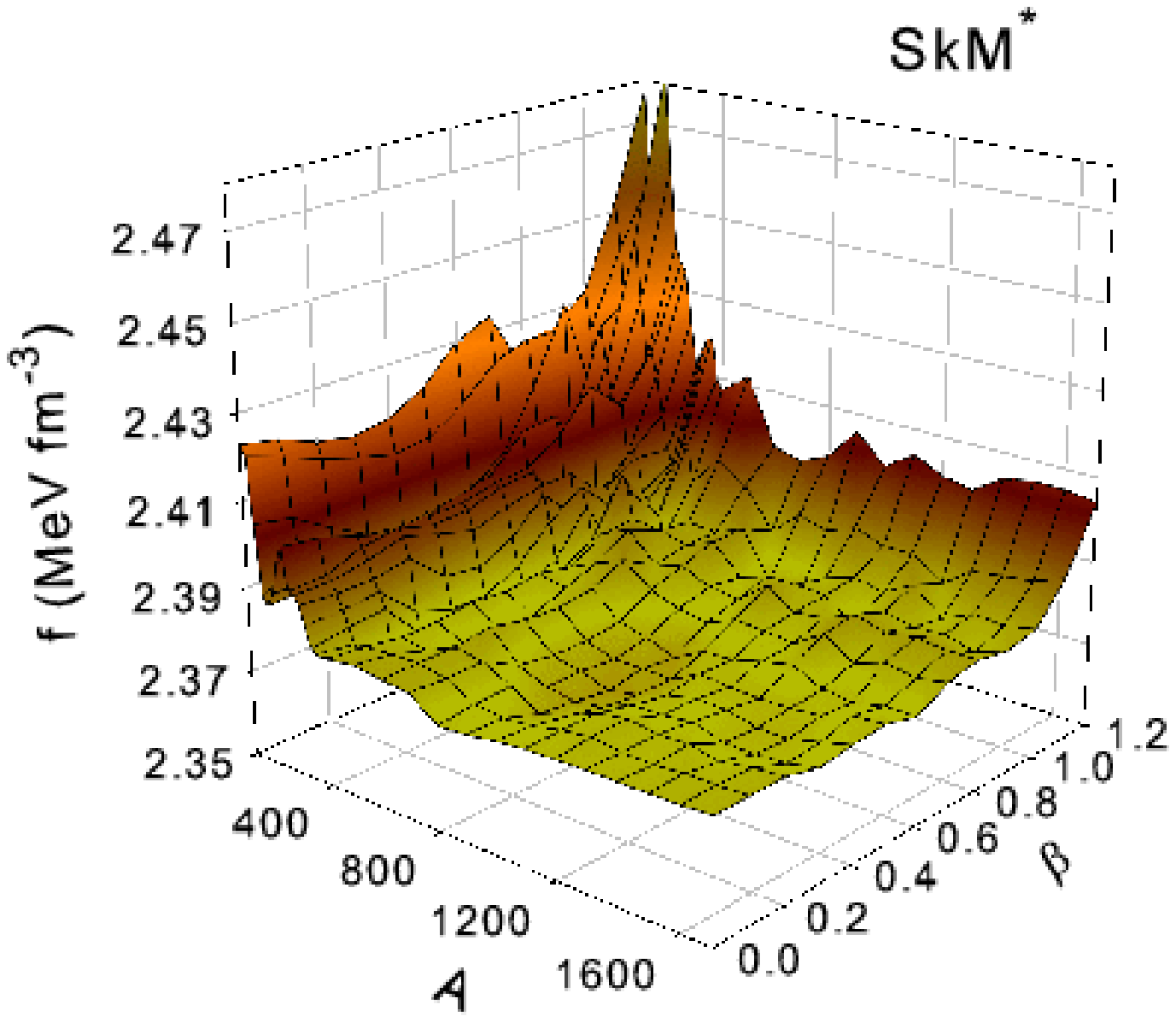,width=7.7cm}\psfig{file=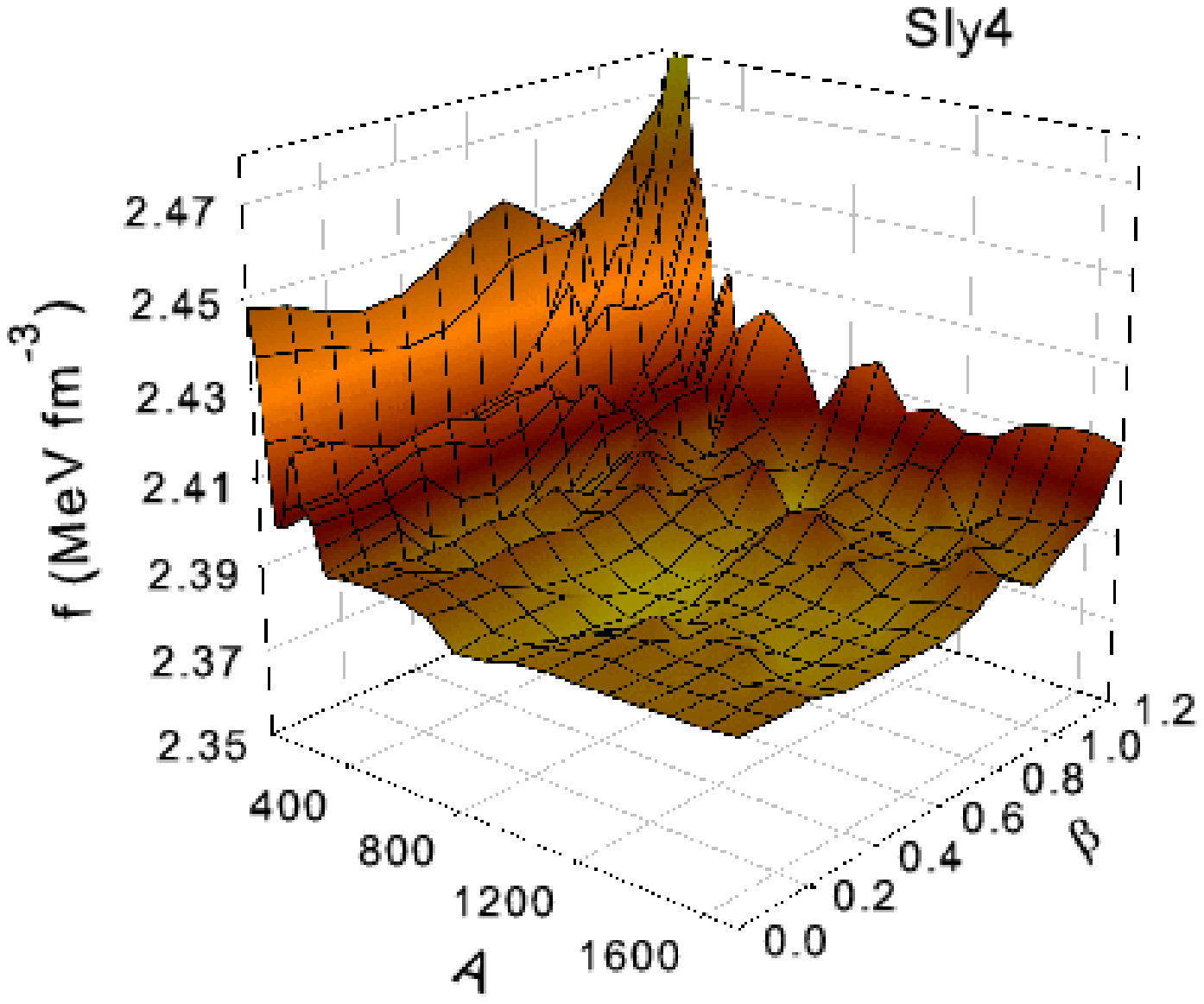,width=7.7cm}}
\caption{(Color on-line) Energy-deformation surfaces at n$_{\rm b}$ = 0.08 fm$^{\rm -3}$, $\gamma$ = $0^{\rm o}$ and $T = 0.0$ MeV as calculated for two different Skyrme parameterizations SkM$^*$ (left) and Sly4 (right).} \hspace{1pc} \label{Fig:16}
\end{figure}

In order to examine the dependence of the energy-deformation surfaces, and ultimately the pasta phase diagram, on the properties of uniform nuclear matter (e.g. compressibility, symmetry energy), the results of calculations using different nuclear interactions should be compared. Fig.~\ref{Fig:16} shows plots of the energy-deformation surface at $n_{\rm b}$ = 0.08 fm$^{\rm -3}$, $\gamma$ = $0^{\rm o}$ and $T = 0.0$ MeV as calculated for two different Skyrme parameterizations SkM$^*$ and Sly4. The surfaces are qualitatively similar, with the Sly4 force predicting a systematically higher free energy-density than SkM$^*$ by $\sim 10$ keV fm$^{\rm -3}$. Sly4 has a higher symmetry energy than SkM$^*$ throughout the density and temperature region under investigation, and thus the energy cost of adding protons to the matter is higher. A broad minimum in $f$ occurs for both parameterizations at $A \approx 1000$, $\beta \approx 0.5$ corresponding to a cylindrical hole configuration; the free energy-density for that minimum is $f$ = 2.3709 MeV fm$^{-3}$ for SkM$^*$ and $f$ = 2.3816 MeV fm$^{-3}$ for Sly4. A more sharply defined minimum occurs for each parameterization at $A = 800$, $\beta = 1.0$ corresponding to a slab-like nuclear shape; there, $f$ = 2.3722 MeV fm$^{-3}$ for SkM$^*$ and $f$ = 2.3803 MeV fm$^{-3}$ Sly4. For both Skyrme parameterizations these constitute the two lowest energy local minima, although the lower of the two differs between the two parameterizations. The difference in energy-density between the minima is $\sim 1$ keV fm$^{\rm -3}$ in both cases. We can conclude that the two parameterizations offer a qualitatively similar spectrum of local minima at this particular density and temperature. A more detailed comparison at a variety of densities and temperatures should be carried out and extended to a wider range of Skyrme parameterizations (and other compatible energy-density functionals) to test the dependence of the pasta phase diagram on properties of the EoS of uniform nuclear matter.

\section{\label{sec4}Summary and conclusions}

The self-consistent 3D Skyrme-Hartree-Fock+BCS method has been applied to the study of the inhomogeneous `pasta' phase of bulk nuclear matter through the use of a new code TAMAR. A constraint was imposed on both independent quadrupole moments of the neutron density. Constraining them to a constant value and varying the cell size allows us to calculate the variation in free energy-density for a specific nuclear shape as its size (or, correspondingly, the number of nucleons it comprises) increases. Variation of the quadrupole moments then allows a self-consistent determination of the energy-density of different nuclear geometries at a given density and temperature. The variation of all free parameters at a given density and temperature plots out an energy-deformation surface $f = f(A, \beta, \gamma)$. In this paper the code was applied to the calculation of the properties of inhomogeneous nuclear matter in a core-collapse supernova with a fixed value of the proton fraction $y_{\rm p} = 0.3$.

It has been demonstrated that the present model smoothly spans the transition from isolated, roughly spherical nuclei surrounded by a low density gas of nucleons to uniform nuclear matter with increasing temperature and baryon number density. The free energy-density of non-uniform matter smoothly tends toward that of uniform matter as the transition density is reached. Discontinuities in the pressure and entropy are suggestive of a phase transition to uniform matter; further calculations are required to determine whether that phase transition is first or second order. The five canonical types of pasta are seen to emerge naturally from the model as well as a large variety of nuclear shapes intermediate between them. At a given density and temperature, many local minima occur corresponding to various nuclear geometries, indicating that a variety of pasta shapes coexist; as in neutron star matter \cite{magie02} the pasta phase in supernova matter is likely to be somewhat disordered. This leads to the suggestion that the thermodynamical properties vary smoothly throughout the pasta regime rather than a series of phase transitions occurring from one bulk phase made up of one specific pasta shape to another: the properties of the different pasta configurations that make up the matter are likely to be averaged on a meso- and macroscopic scale.

Special attention was paid to accounting for all the possible numerical effects that might contribute to the results. The choice of original wavefunctions used to start the iteration solution to the HF equations and the finite size of the computational cell were shown to have no effect on the nuclear geometries obtained. Spurious (numerical) shell effects, arising from the artificial discretization of continuum neutron states by the finite computational cell have been identified. These effects give rise to oscillatory variation in the energy-density of matter with the computational cell size that is distinct from the physical shell effects, allowing easy identification when they are the dominant shell energy present. At densities just below the density of transition to uniform matter, where the spurious shell effects most strongly obscure the physical minima in the energy surfaces, their contribution to the total energy-density of matter is in the range $\sim 10-50$ keV fm$^{-3}$ at $A \lesssim 400$. In order to obtain the correct physical energy minima and the transition density with accuracy, the spurious shell effects must be taken into account. One possibility is to use analytic expressions for the artificial shell energy obtained using the semi-classical WKB approximation~\cite{SC1}; this will be investigated in a subsequent paper.

The plethora of local minima in the energy-deformation surfaces $f(A, \beta, \gamma)$ arise from a combination of physical shell effects from bound nucleons and a number of broad `geometrical' minima occurring over different regions of the energy surface that arise from the competition between the surface energy and Coulomb energies of the different nuclear shapes. Physical shell effects that arise from the interaction of the continuum neutrons with the nuclear structures, expected to be a significant contribution to the energy spectrum in neutron star crustal matter \cite{magie02}, play a negligible role in supernova matter as the neutron gas external to the nuclear structures is generally an order of magnitude less dense than in supernova matter (strongly linked to the fact that the proton fraction is roughly order of magnitude higher in supernova matter). Geometrical minima are separated from each other by energy barriers in the energy-deformation surfaces of the order 1 - 5 keV fm$^{-3}$. At low temperatures $T \le 2.5$ MeV, physical shell effects contribute similarly deep local minima. The shell effects decrease in magnitude as temperature increases, vanishing at $T$ $\approx 5$ MeV and above.

The energy-deformation surface at $n_{\rm b}$ = 0.08 fm$^{\rm -3}$, $\gamma$ = $0^{\rm o}$ and $T = 0.0$ MeV as calculated with the SkM$^*$ Skyrme parameterization was compared to the same surface calculated by the Sly4 parameterization. The surfaces were qualitatively similar, with local minima appearing in the same regions of parameter space corresponding to the same nuclear shapes. A more detailed examination of the dependence of the pasta phase diagram on the EoS of uniform nuclear matter is underway.

Work is in progress on extending of the present calculation over a larger area of baryon number density, proton fraction and temperature in order to construct EoS table for modeling of core-collapse supernovae and pinpoint more accurately the position in density-temperature space and the nature of the phase transition to uniform matter.

\section*{Acknowledgement}
The authors gratefully acknowledge Tony Mezzacappa for stimulating discussions, encouragement and support of access to the leadership computing facility in Oak Ridge, where majority of the calculations were done, and to NERSC computers at Lawrence Berkeley National Laboratory used during the initial phase. It is a pleasure to thank John Miller for continuous encouragment and comments during the course for this work, and to Chris Pethick and Philipp Podsiadlowski for interesting discussions and suggestions. Thanks also to P.-G. Reinhard, Raph Hix, Bronson Messer and N.J.Stone for helpful comments and to Constanca Providencia and Miguel Oliveira for supporting the final stage of the calculations at the super-computer Milipeia at Universidade de Coimbra, Portugal. An important part of the project was the visualization of the results which would not have been possible without substantial help of Chaoli Wang, Jonathan Edge, Amy Bonsor and Ross Toedte.

The major part of this work was conducted under the auspices of the TeraScale Supernova Initiative, funded by SciDAC grants from the DOE Office of Science High-Energy, Nuclear, and Advanced Scientific Computing Research Programs. Resources of the Center for Computational Sciences at Oak Ridge National Laboratory were used. Oak Ridge National Laboratory is managed by UT-Battelle, LLC, for the U.S. Department of Energy under contract DE-AC05-00OR22725. Partial support for this work by the UK EPSRC (WGN) and US DOE grant DE-FG02-94ER40834 (JRS) is acknowledged with thanks.

\newpage

\begin{figure}[!t]
\caption{(Color) Obtaining a double nuclear shape. The right picture shows a 3D rendering of the neutron density profile at $n_{\rm b}$ = 0.08 fm$^{\rm -3}$, $y_{\rm p}$ = 0.3, $T$ = 0.0 MeV, $A$ = 700, and $(\beta,\gamma)$ = (1.0, 0$^{\rm o}$). The configuration shown in the left picture is the result of a calculation in a box double the size in the z-direction - with $A$ = 1400 - at otherwise the same parameter values. Blue indicates the lowest densities and red the highest.} \hspace{1pc} \label{Fig:1}
\end{figure}

\newpage
\clearpage

\newpage
\clearpage

\newpage
\clearpage

\begin{figure}[!t]
\caption{(Color) 3D renderings of the neutron density profiles of configurations taken from the sequences of increasing $A$ starting at $A$ = 100 and increasing in steps of 100 from left to right and from top to bottom (snake fill) up until the last row, where the configurations that deviate from the main nuclear geometry are shown (see text for details). Blue indicates the lowest densities and red the highest.} \label{Fig:5}
\end{figure}

\newpage
\clearpage

\newpage
\clearpage

\newpage
\clearpage

\newpage
\clearpage

\begin{figure}[!t]
\hspace{1pc}
\caption{(Color) 3D renderings of the neutron density profiles of minimum energy configurations at $T$ = 2.5 MeV and densities of $n_{\rm b}$ = 0.04 fm$^{-3}$ (top left), $n_{\rm b}$ = 0.06 fm$^{-3}$ (top middle), $n_{\rm b}$ = 0.08 fm$^{-3}$ (top right), $n_{\rm b}$ = 0.09 fm$^{-3}$ (bottom left), $n_{\rm b}$ = 0.10 fm$^{-3}$ (bottom middle) and $n_{\rm b}$ = 0.11 fm$^{-3}$ (bottom right). Blue indicates the lowest densities and red the highest. } \hspace{1pc} \label{Fig:11}
\end{figure}

\begin{figure}[!b]
\hspace{1pc}
\caption{(Color) 3D renderings of the neutron density profiles of minimum energy configurations at $n_{\rm b}$ = 0.10 fm$^{-3}$ and temperatures of $T$ = 0 MeV (top left), 2.5 MeV (top right), 5 MeV (bottom left) and 7.5 MeV (bottom right). Blue indicates the lowest densities and red the highest.} \hspace{1pc} \label{Fig:12}
\end{figure}

\newpage
\clearpage

\newpage
\clearpage

\newpage
\clearpage

\newpage
\clearpage

\begin{table}[!t]
    \begin{center}
        \caption{Comparison of nuclear binding energies and root mean square radii of nuclei for three different codes using the Sly4 Skyrme parametrization. Coulomb and spin-orbit potentials were omitted in the calculation. See text for details of the three codes.} \label{Tab:1}
        \vspace{0.2cm}
        \begin{tabular}{p{2cm} p{1.8cm} p{1.8cm} p{1.8cm} p{2.5cm} p{2.5cm} p{2.5cm}} \hline \rule[-8pt]{0pt}{22pt}
            & \multicolumn{3}{p{4.5cm}}{Binding Energy (MeV)}  & \multicolumn{3}{p{7.5cm}}{Proton (Neutron) RMS Radius (fm)}  \\ \hline
            \hspace{0.5cm}Nucleus & SKYAX & TDHF3D & TAMAR & SKYAX & TDHF3D & TAMAR \\ \hline
            \hspace{0.5cm} $^{16}$O & 126.81 & 126.80 & 126.80 & 2.701 (2.701) & 2.701 (2.701) & 2.701 (2.701) \\
            \hspace{0.5cm}$^{40}$Ca & 399.49 & 399.33 & 399.35 & 3.374 (3.374) & 3.373 (3.373) & 3.373 (3.373) \\
            \hspace{0.5cm}$^{56}$Fe & 564.13 & 563.96 & 563.97 & 3.789 (3.875) & 3.789 (3.877) & 3.790 (3.878) \\ \hline
        \end{tabular}
    \end{center}
\end{table}

\begin{table}[!b]
    \begin{center}
        \caption{EoS Quantities vs A for $n_{\rm b}$ = 0.06 fm$^{-3}$, T = 2.5 MeV, $\beta, \gamma$ = (1.0, 0$^{\rm o}$). Free energy-density $f$ (MeV fm$^{-3}$), entropy density $s$ (MeV fm$^{-3}$) and pressure density $p$ (MeV fm$^{-3}$) are given.} \label{Tab:2}
        \vspace{0.2cm}
        \begin{tabular}{p{2cm} p{2cm} p{2cm} p{1.3cm}} \hline \rule[-8pt]{0pt}{22pt}
        \hspace{0.2cm} $A$ & $f$ & $s$ & $p$ \\ \hline
        \hspace{0.2cm} 200 & 1.5603 & 0.0907 & 0.6601 \\
        \hspace{0.2cm} 300 & 1.5529 & 0.0891 & 0.6571 \\
        \hspace{0.2cm} 400 & 1.5487 & 0.0891 & 0.6623 \\
        \hspace{0.2cm} 500 & 1.5464 & 0.0873 & 0.6662 \\
        \hspace{0.2cm} 600 & 1.5456 & 0.0865 & 0.6661 \\
        \hspace{0.2cm} 700 & 1.5455 & 0.0868 & 0.6635 \\
        \hspace{0.2cm} 800 & 1.5457 & 0.0870 & 0.6614 \\
        \hspace{0.2cm} 900 & 1.5460 & 0.0867 & 0.6593 \\
        \hspace{0.2cm} 1000 & 1.5464 & 0.0862 & 0.6573 \\
        \hspace{0.2cm} 1100 & 1.5470 & 0.0859 & 0.6548 \\
        \hspace{0.2cm} 1200 & 1.5478 & 0.0859 & 0.6523 \\ \hline
        \end{tabular}
    \end{center}
\end{table}

\newpage
\clearpage

\begin{table}[!t]
    \begin{center}
        \caption{EoS at $T$ = 0 MeV. Quantities given are: baryon number density $n_{\rm b}$ (fm$^{-3}$), nucleon number $A$ and deformation parameters $\beta$, $\gamma$ for minimum energy configuration, entropy density $s$ (MeV fm$^{-3}$), free energy-density $f$ (MeV fm$^{-3}$), pressure $p$ (MeV fm$^{-3}$), proton, neutron and electron chemical potentials $\mu_{\rm p}$ (MeV), $\mu_{\rm n}$ (MeV), $\mu_{\rm e}$ (MeV), and $r_{\rm RMS}$ divided by half the cell width, $\delta$.} \label{Tab:3}
        \vspace{0.2cm}
        \begin{tabular}{p{1.3cm} p{1.3cm} p{1.3cm} p{1.3cm} p{1.3cm} p{1.3cm} p{1.3cm} p{1.3cm} p{1.3cm} p{1.3cm} p{0.9cm}} \hline \rule[-8pt]{0pt}{22pt}
        \hspace{0.2cm} $n_{\rm b}$  & A & $\beta$ & $\gamma$ & $s$ & $f$ & $p$ & $\mu_{\rm p}$ & $\mu_{\rm n}$ & $\mu_{\rm e}$ & $\delta$ \\ \hline
        \hspace{0.2cm} 0.04 & 500 & 1.3 & 30$^{\rm o}$ & 0.0   & 0.901 & 0.370 & -32.4 & -1.27 & 139.3 & 0.71 \\
        \hspace{0.2cm} 0.06 & 320 & 1.4 & 0$^{\rm o}$  & 0.0   & 1.593 & 0.609 & -35.4 & -1.62 & 149.5 & 0.88 \\
        \hspace{0.2cm} 0.08 & 400 & 0.9 & 60$^{\rm o}$ & 0.0   & 2.364 & 0.906 & -37.4 & -1.11 & 176.4 & 0.88 \\
        \hspace{0.2cm} 0.09 & 1000& 0.3 & 0$^{\rm o}$  & 0.0   & 2.786 & 1.019 & -39.6 & -1.80 & 182.6 & 0.90 \\
        \hspace{0.2cm} 0.10 & 1100& 0.0 & 0$^{\rm o}$  & 0.0   & 3.220 & 1.146 & -40.6 & -2.19 & 189.2 & 0.93 \\
        \hspace{0.2cm} 0.11 & 1300& 0.0 & 0$^{\rm o}$  & 0.0   & 3.662 & 1.152 & -42.3 & -3.95 & 195.3 & 1.00 \\
        \hspace{0.2cm} 0.12 & 1200& 0.0 & 0$^{\rm o}$  & 0.0   & 4.124 & 1.447 & -39.7 & -2.59 & 201.1 & 1.00 \\ \hline
        \end{tabular}
    \end{center}
\end{table}

\begin{table}[!b]
    \begin{center}
        \caption{EoS at $T$ = 2.5 MeV. Quantities as in Table~\ref{Tab:3}.} \label{Tab:4}
        \vspace{0.2cm}
        \begin{tabular}{p{1.3cm} p{1.3cm} p{1.3cm} p{1.3cm} p{1.3cm} p{1.3cm} p{1.3cm} p{1.3cm} p{1.3cm} p{1.3cm} p{0.9cm}} \hline \rule[-8pt]{0pt}{22pt}
        \hspace{0.2cm} $n_{\rm b}$ & A & $\beta$ & $\gamma$ & $s$ & $f$ & $p$ & $\mu_{\rm p}$ & $\mu_{\rm n}$ & $\mu_{\rm e}$ & $\delta$ \\ \hline
        \hspace{0.2cm} 0.04 & 400 & 1.6 & 0$^{\rm o}$  & 0.064 & 0.863 & 0.393 & -31.9 & -1.89 & 139.1 & 0.73 \\
        \hspace{0.2cm} 0.06 & 340 & 1.3 & 60$^{\rm o}$ & 0.086 & 1.533 & 0.654 & -35.5 & -1.90 & 159.3 & 0.84 \\
        \hspace{0.2cm} 0.08 & 800 & 0.8 & 60$^{\rm o}$ & 0.106 & 2.305 & 0.947 & -37.9 & -1.77 & 175.5 & 0.88 \\
        \hspace{0.2cm} 0.09 & 1000& 0.2 & 0$^{\rm o}$  & 0.115 & 2.718 & 1.060 & -39.6 & -2.10 & 182.5 & 0.90 \\
        \hspace{0.2cm} 0.10 & 1100& 0.0 & 0$^{\rm o}$  & 0.124 & 3.145 & 1.198 & -41.1 & -2.27 & 189.1 & 0.93 \\
        \hspace{0.2cm} 0.11 & 1300& 0.0 & 0$^{\rm o}$  & 0.134 & 3.583 & 1.299 & -42.7 & -2.84 & 195.2 & 1.00 \\
        \hspace{0.2cm} 0.12 & 1200& 0.0 & 0$^{\rm o}$  & 0.137 & 4.041 & 1.577 & -43.6 & -1.82 & 201.0 & 1.00 \\ \hline
        \end{tabular}
    \end{center}
\end{table}

\newpage
\clearpage

\begin{table}[!t]
    \begin{center}
        \caption{EoS at $T$ = 5 MeV. Quantities as in Table~\ref{Tab:3}.} \label{Tab:5}
        \vspace{0.2cm}
        \begin{tabular}{p{1.3cm} p{1.3cm} p{1.3cm} p{1.3cm} p{1.3cm} p{1.3cm} p{1.3cm} p{1.3cm} p{1.3cm} p{1.3cm} p{0.9cm}} \hline \rule[-8pt]{0pt}{22pt}
        \hspace{0.2cm} $n_{\rm b}$ & A & $\beta$ & $\gamma$ & $s$ & $f$ & $p$ & $\mu_{\rm p}$ & $\mu_{\rm n}$ & $\mu_{\rm e}$ & $\delta$ \\ \hline
        \hspace{0.2cm} 0.04 & 400 & 1.6 & 0$^{\rm o}$  & 0.261 & 0.745 & 0.455 & -31.8 & -3.83 & 138.6 & 0.76 \\
        \hspace{0.2cm} 0.06 & 600 & 1.1 & 60$^{\rm o}$ & 0.356 & 1.373 & 0.741 & -35.6 & -3.42 & 159.0 & 0.85 \\
        \hspace{0.2cm} 0.08 & 800 & 0.4 & 0$^{\rm o}$  & 0.437 & 2.108 & 1.021 & -38.8 & -3.45 & 175.1 & 0.89 \\
        \hspace{0.2cm} 0.09 & 1200& 0.0 & 0$^{\rm o}$  & 0.478 & 2.501 & 1.165 & -40.2 & -3.50 & 182.2 & 0.92 \\
        \hspace{0.2cm} 0.10 & 1200& 0.0 & 0$^{\rm o}$  & 0.515 & 2.909 & 1.327 & -41.6 & -3.41 & 188.7 & 0.96 \\
        \hspace{0.2cm} 0.11 & 1200& 0.0 & 0$^{\rm o}$  & 0.545 & 3.332 & 1.505 & -43.0 & -3.14 & 194.9 & 1.00 \\
        \hspace{0.2cm} 0.12 & 1200& 0.0 & 0$^{\rm o}$  & 0.555 & 3.784 & 1.786 & -43.8 & -1.84 & 200.7 & 1.00 \\ \hline
        \end{tabular}
    \end{center}
\end{table}


\begin{thebibliography}{99}
\bibitem{Liebendorfer2008}
M. Liebend\"orfer, T.Fischer, C.Fr\"ohlich, W.R.Hix, K.Langanke, G.Martinez-Pinedo, A.Mezzacappa, S.Scheidegger, F.-K.Thielemann, and S.C.Whitehouse, New Astron. Reviews. \textbf{52}, 373 (2008)
\bibitem{gnedin01}
O.Y. Gnedin, D.G. Yakovlev, and A.Y. Potekhin, Mon. Not. R. Astron. Soc. \textbf{324}, 725 (2001)
\bibitem{page04}
D. Page, J.M. Lattimer, M. Prakash, A.W. Steiner, Astrophys. J. Suppl. Ser. \textbf{155}, 623 (2004)
\bibitem{Yak2008}
D.G. Yakovlev, O.Y. Gnedin, A.D. Kaminker and A.Y. Potekhin, AIP Conference Series \textbf{983}, 379 (2008)
\bibitem{Andersson2002}
N. Andersson, G.L. Comer, D. Langlois, Phys. Rev. \textbf{D66}, 104002 (2002)
\bibitem{Samuelsson2007}
L. Samuelsson and N. Andersson, Mon. Not. R. Astron. Soc. \textbf{374}, 256 (2007)
\bibitem{horva04}
J.A. Horvath, Int.J.Mod.Phys. \textbf{D13}, 1327, (2004)
\bibitem{larso02}
M.B. Larson and B. Link, Mon. Not. R. Astron. Soc. \textbf{333} 613, (2002)
\bibitem{crawf03}
F. Crawford and M. Demianski, Ap.J. \textbf{595} 1052, (2003)
\bibitem{jones98}
P.B. Jones, Mon. Not. R. Astron. Soc. \textbf{296} 217, (1998)
\bibitem{Lin2007}
L-M. Lin, N. Andersson, G.L. Comer, Phys. Rev. \textbf{D78}, 083008 (2008)
\bibitem{peth95}
C.J. Pethick, D.G. Ravenhall, C.P. Lorenz, Nucl. Phys. \textbf{A584} 675, (1995)
\bibitem{lattimer78}
J.M. Lattimer and D.G. Ravenhall, ApJ \textbf{223} 314, (1978)
\bibitem{lattimer81}
J.M. Lattimer, Abb. Rev. Nucl. Part. Sci. \textbf{31} 337, (1981)
\bibitem{lattimer91}
J.M. Lattimer and F.Douglas Swesty, Nucl. Phys. \textbf{A535} 331, (1991)
\bibitem{Freedman74}
D.Z. Freedman, Phys. Rev. \textbf{D9}, 1389 (1974)
\bibitem{Sato75}
K.Sato, Prog. Theor. Phys. \textbf{53}, 595 (1975)
\bibitem{rave83}
D.G. Ravenhall, C.J. Pethick, J.R. Wilson, PRL \textbf{50}, 26, 2066, (1983)
\bibitem{hashi84}
M. Hashimoto, H. Seki, M. Yamada, Prog. Th. Phys. \textbf{71}, 2, 320, (1984)
\bibitem{hashi84_2}
K. Oyamatsu, M. Hashimoto, M. Yamada, Prog. Th. Phys. \textbf{72} 2 373, (1984)
\bibitem{watanabe05}
G. Watanabe, H. Sonoda, \emph{Soft Condensed Matter: New Research} (Nova Science Publishers 2007), arXiv:cond-mat/0502515
\bibitem{sonoda07}
H. Sonoda, G. Watanabe, K. Sato, T. Takiwaki, K. Yasuoka, T. Ebisuzaki, Phys. Rev. \textbf{C75}, 042801(R), (2007)
\bibitem{sonoda08}
H. Sonoda, G. Watanabe, K. Sato, K. Yasuoka, T. Ebisuzaki, Phys. Rev. \textbf{C77}, 035806 (2008)
\bibitem{lassaut87}
M. Lassaut, H. Flocard, P.Bonche, P.H.Heenen, E.Suraud, Astron. Astrophys. \textbf{183}, L3 (1987)
\bibitem{bonche81}
P. Bonche and D. Vautherin, Nucl. Phys. \textbf{A372} 496, (1981)
\bibitem{bonche82}
P. Bonche and D. Vautherin, A\&A \textbf{112} 268, (1982)
\bibitem{maruy1998}
T. Maruyama, K. Niita, K. Oyamatsu, T. Maruyama, S. Chiba, A. Iwamoto, Phys. Rev. \textbf{C57} 655, (1998)
\bibitem{horowitz2004_1}
C.J.Horowitz, M.A. Perez-Garcia, J.Piekarewicz, Phys. Rev. \textbf{C69}, 045804, (2004)
\bibitem{horowitz2004_2}
C.J.Horowitz, M.A. Perez-Garcia, J.Carriere, D.K.Berry, J.Piekarewicz, Phys. Rev. \textbf{C70}, 065806, (2004)
\bibitem{watanabe01_2}
G. Watanabe, K. Iida, K. Sato, Prog. Th. Phys. \textbf{106} 551, (2001)
\bibitem{watanabe_2_05}
G. Watanabe, H. Sonoda, \emph{Reaction Mechanisms for Isotope Beams}, AIP Conference Series \textbf{791}, 101, (2005)
\bibitem{magie02}
P. Magierski, P.-H. Heenen, Phys. Rev. \textbf{C65}, 045804, (2002)
\bibitem{gogel07}
P. G\"ogelein, E.N.E. van Dalen, C. Fuchs, H. M\"uther, Phys. Rev. \textbf{C77}, 025802, (2008)
\bibitem{magie03}
P. Magierski, A. Bulgac, Nucl. Phys. \textbf{A719} 217, (2003)
\bibitem{Skyrme1956}
T.H.R. Skyrme, Phil. Mag. \textbf{1} 1043 (1956)
\bibitem{Vaut1972}
D. Vautherin, D.M. Brink, Phys. Rev. \textbf{C5} 3 626, (1972)
\bibitem{Langa91}
K. Langanke, J.A. Maruhn, S.E. Koonin, \emph{Computational Nuclear Physics 1},
Springer-Verlag, New York, (1991).
\bibitem{sto07}
J.R. Stone and P.G. Reinhard, Prog.Part.Nucl.Phys. \textbf{58}, 587 (2007)
\bibitem{new06}
W.G.Newton, J.R.Stone, T. Mezzacappa, J.Phys.CS \textbf{46} 408, (2006)
\bibitem{Bonche1987}
P. Bonche, H. Flocard, P.H. Heenen, Nucl. Phys. \textbf{A467}, 1, 115 (1987)
\bibitem{Bender2003}
M. Bender, P.H.Heenen and P.G.Reinhard, Rev. Mod. Phys. \textbf{75}, 1, 121 (2003)
\bibitem{thesis}
W.G.Newton, DPhil Thesis, University of Oxford (2008),
http://www.tamu-commerce.edu/home/physics/newton/thesis.pdf
\bibitem{Bartel1982}
J. Bartel, P. Quentin, M. Brack, C. Guet, H.-B. Hakansson, Nucl. Phys. \textbf{A386}, 79 (1982)
\bibitem{Chaba1998}
E. Chabanat, P. Bonche, P. Haensel, J. Meyer, R. Schaeffer, Nucl. Phys. \textbf{A635}, 231 (1998)
\bibitem{stone03}
J.R. Stone, J.C. Miller, R. Koncewicz, P.D. Stevenson, M.R. Strayer, Phys. Rev. \textbf{C68}, 3, 034324 (2003)
\bibitem{Cuss1985}
R.Y. Cusson, P.G. Reinhard, M.R. Strayer, J.A. Maruhn, W. Greiner, Z. Phys. A - Atoms and Nuclei \textbf{320}, 475 (1985)
\bibitem{Bonche1985}
P. Bonche, H. Flocard, P.H. Heenen, S.J.Krieger, M.S. Weiss, Nucl. Phys. \textbf{A443}, 1, 39 (1985)
\bibitem{fftw}
http://www.fftw.org/
\bibitem{bloch}
N.W. Ashcroft and N.D. Mermin, \emph{Solid State Physics}, (Philadelphia: Saunders College 1976).
\bibitem{carter05}
B. Carter, N. Chamel, P. Haensel, Nucl.Phys. \textbf{A748} 675, (2005)
\bibitem{slater1951}
J.C. Slater, 1951, Phys. Rev. \textbf{81}, 385 (1951)
\bibitem{Baye1986}
D. Baye and P.-H. Heenen, J.Phys. \textbf{A19}, 2041 (1986)
\bibitem{Davi1980}
K.T.R. Davies, H. Flocard, S. Krieger, M.S. Weiss, Nucl. Phys. \textbf{A342}, 113 (1980).
\bibitem{Rein1982}
P.-G. Reinhard and R.Y. Cusson, Nucl. Phys. \textbf{A378}, 418 (1982)
\bibitem{Umar1989}
A.S. Umar, M.R. Strayer, P.-G. Reinhard, K.T.R. Davies, S.-J. Lee, Phys.Rev. \textbf{C40}, 706 (1989)
\bibitem{rein_private2}
P.-G. Reinhard, \emph{private communication} (2007)
\bibitem{SC1}
M. Brack and R.K. Bhaduri, \emph{Semiclassical physics}, Addison-Wesley (1997).
\end{thebibliography}
\end{document}